\documentclass[nocrop]{myClass}
\copyrightyear{2016} \pubyear{?}

\usepackage{natbib}
\usepackage{amssymb}
\usepackage{amsmath}
\usepackage{multirow}

\renewcommand\tilde\widetilde 
\renewcommand\hat\widehat 
\renewcommand\epsilon\varepsilon 
\newcommand{\tr}[1]{\ensuremath{#1^{T}}} 
\newcommand{\mb}[1]{\ensuremath{\mathbf{#1}}} 
\newcommand{\mbg}[1]{\ensuremath{\boldsymbol{#1}}} 

\newcommand\argmin{\text{argmin}}

\newcommand\Cov{\text{Cov}}
\newcommand\EE{\mathbb{E}} 
\newcommand\PP{\mathbb{P}}
\newcommand\RR{\mathbb{R}} 
\newcommand\II{\mathbf{1}}
\newcommand{\btimes}{\mathbin{\protect\rotatebox[origin=c]{45}{\scriptsize$\boxtimes$}}}
\newcommand{\diam}{\mathbin{\protect\rotatebox[origin=c]{45}{\tiny$\blacksquare$}}}
\newcommand\msum{{\textstyle\sum}}
\newcommand\like{\mathcal{L}}

\access{}
\appnotes{}

\begin{document}
\firstpage{1}

\subtitle{}

\title[Adaptive Sparse PLS for Logistic Regression]{High Dimensional Classification with combined Adaptive Sparse PLS and Logistic Regression}
\author[G.~Durif \textit{et~al}.]{G.~Durif\,$^{\text{\sfb 1,\sfb 2,}*}$, L.~Modolo\,$^{\text{\sfb 1,\sfb 3,\sfb 4}}$, J. Michaelsson\,$^{\text{\sfb 4}}$, J. E. Mold\,$^{\text{\sfb 4}}$, S.~Lambert-Lacroix\,$^{\text{\sfb 5}}$ and F.~Picard\,$^{\text{\sfb 1}}$}
\address{$^{\text{\sf 1}}$LBBE, UMR CNRS 5558, Universit\'e Lyon 1, F-69622 Villeurbanne, France, \\
$^{\text{\sf 2}}$INRIA Grenoble Alpes, THOTH team, F-38330 Montbonnot, France, \\
$^{\text{\sf 3}}$LBMC UMR 5239 CNRS/ENS Lyon, F-69007 Lyon, France,\\
$^{\text{\sf 4}}$Department of Cell and Molecular Biology, Karolinska Institutet, Stockholm, Sweden, \\
$^{\text{\sf 5}}$UMR 5525 Universit\'e Grenoble Alpes/CNRS/TIMC-IMAG, F-38041 Grenoble, France.}

\corresp{$^\ast$To whom correspondence should be addressed.}

\history{}

\editor{}

\abstract{\textbf{Motivation:} The high dimensionality of genomic data calls for the development of specific classification methodologies, especially to prevent over-optimistic predictions. This challenge can be tackled  by compression and variable selection, which combined constitute a powerful framework for classification, as well as data visualization and interpretation. However, current proposed combinations lead to unstable and non convergent methods due to inappropriate computational frameworks. We hereby propose a computationally stable and convergent approach for classification in high dimensional based on sparse Partial Least Squares (sparse PLS).  \\
\textbf{Results:} We start by proposing a new solution for the sparse PLS problem that is based on proximal operators for the case of univariate responses. Then we develop an adaptive version of the sparse PLS for classification, called logit-SPLS, which combines iterative optimization of logistic regression and sparse PLS to ensure computational convergence and stability. Our results are confirmed on synthetic and experimental data. In particular we show how crucial convergence and stability can be when cross-validation is involved for calibration purposes.
Using gene expression data we explore the prediction of breast cancer relapse. We also propose a multicategorial version of our method, used to predict cell-types based on single-cell expression data.\\
\textbf{Availability:} Our approach is implemented in the \texttt{plsgenomics} R-package. \\ 
\textbf{Contact:} \href{ghislain.durif@inria.fr}{ghislain.durif@inria.fr}\\
\textbf{Supplementary information:} Supplementary materials are available at \textit{Bioinformatics}
online.}

\maketitle

\section{Introduction}
\label{sec:intro}

Molecular classification is at the core of many recent studies based on Next-Generation Sequencing data. For instance, the genomic characterization of diseases based on genomic signatures has been one \textit{Grail} for many studies to predict patient outcome, survival or relapse \citep{guedj2012}. Moreover, following the recent advances of sequencing technologies, it is now possible to isolate and sequence the genetic material from a single cell \citep{stegle2015}. Single-cell data give the opportunity to characterize the genomic diversity between the individual cells of a specific population. However, in both cases, the specific context of high dimensionality constitutes a major challenge for the development of new statistical methodologies \citep{marimont1979,donoho2000}. Indeed, the number of recorded variables $p$ (as gene expression) being far larger than the sample size $n$, classical regression or classification methods are inappropriate \citep{aggarwal2001,hastie2009}, due to spurious dependencies between variables, that lead to singularities in the optimization processes, with neither unique nor stable solution.

This challenge calls for the development of specific statistical tools, such as the following dimension reduction approaches: $(i)$ Compression methods that search for a representation of the data in lower dimensional space and $(ii$) Variable selection methods, based on a parsimony hypothesis, i.e., among all recorded variables, a lot are supposed to be uninformative and can be considered as noise to be removed from the model. For instance, the Partial Least Squares (PLS) regression \citep{wold1975,wold1983} is a compression approach appropriate for linear regression, especially with highly correlated covariates, that constructs new components, i.e. latent directions, explaining the response. An example of sparsity-based approach is the Lasso \citep{tibshirani1996} where coefficients of less relevant variables are shrunk to zero thanks to a $\ell_1$ penalty in the optimization procedure. Eventually, sparse PLS (SPLS) regression \citep{lecao2008, chun2010} combines both compression and variable selection to reduce dimension. It introduces a selection step based on the Lasso in the PLS framework, constructing new components as sparse linear combinations of predictors. It occurs as well that combining compression and ``sparse'' approach improves the efficiency of prediction and the accuracy of selection. Such an association (compression and selection) is also relevant for data visualization, a crucial challenge when considering high dimensional data. Existing SPLS methods are based on resolutions of approximations of the associated optimization problem. In this work, we first propose a new formulation of the sparse PLS optimization problem with a simple exact resolution, derived from proximal operators \citep{bach2012}. We also introduce an adaptive sparsity-inducing penalty, inspired from the adaptive Lasso \citep{zou2006a}, to improve the variable selection accuracy.

SPLS has shown excellent performance for regression with a continuous response, but its adaptation to classification is not straightforward. \cite{chung2010} or \cite{lecao2011} proposed to use sparse PLS as a preliminary dimension reduction step before a standard classification method, such as discriminant analysis (SPLS-DA) or logistic regression, following previous approaches using classical PLS for molecular classification \citep{nguyen2002, boulesteix2004}. Their approach gives interesting results in SNPs data analysis \citep{lecao2011} or in tumor classification \citep{chung2010}.

Another method for classification consists in using logistic regression (binary or multicategorial) \citep{mccullagh1989}, for which optimization is achieved via the Iteratively Reweighted Least Squares (IRLS) algorithm \citep{green1984}. However its convergence is not guaranteed especially in the high dimensional case. Computational convergence is a crucial issue when estimating parameters, as non-convergent methods may lead to unstable and inconsistent estimations, impacting analysis interpretation and reproducibility, especially when tuning hyper-parameters by cross-validation.

The combination of logistic regression and (sparse) PLS could lead to a classification method processing dimension reduction based on lower space representation and variable selection. However, the combination of such iterative algorithms is not necessarily straightforward, due to convergence issues. Performing compression with SPLS on the categorical response as a first step before logistic regression remains counter-intuitive, because SPLS was designed to handle a continuous response within homoskedastic models. Based on the generalized PLS by \cite{marx1996} or \cite{ding2005}, \cite{chung2010} proposed to use sparse PLS within the IRLS iterations to solve reweighted least squares at each step, however we will see that convergence issues remain. \cite{fort2005} proposed to use a Ridge regularization \citep{eilers2001} to ensure the convergence of the IRLS algorithm and to use the classical PLS to estimate predictor coefficients by using a continuous pseudo-response generated by the IRLS algorithm. We will develop a similar approach based on sparse PLS.

Our new SPLS-based approach, called logit-SPLS, combines compression and variable selection in a GLM framework. We show the accuracy, the computational stability and convergence of our method, compared with other state-of-the-art approaches on simulations. Especially, we show that compression increases variable selection accuracy, and that our method is more stable regarding the choice of hyper-parameters by cross-validation, contrary to other methods processing classification with sparse PLS. Thus, our method is the only one that correctly performs considering all criteria (prediction, selection, stability), whereas all the other approaches present a weak spot. Our simulations illustrate the interest of both selection and compression over selection or compression only. Our work was implemented in the existing R-package \texttt{plsgenomics}, available on the CRAN. 

We will first introduce our adaptive sparse PLS approach. Then, we will develop and discuss our classification framework based on Ridge IRLS and adaptive sparse PLS for logistic regression. We will finish by a comparative study and eventually two applications of our method: $(i)$ binary classification to predict breast cancer relapse after 5 years based on gene expression data, with an illustration of data visualization through compression, $(ii)$ prediction of cell types with multinomial classification based on single-cell expression profiles. To do so, we extend our approach to the multi-group case, based on a ``one-class vs a reference'' type of multi-classification. One strength of our approach is to propose a sparse PLS that admits a closed-form solution in both binary and multi-group classifications. This leads to computationally efficient procedures in both cases, contrary to sparse PLS-DA approaches for instance, that are based on a multivariate response sparse PLS algorithm in the multi-group case, for which there is no closed-form solution \citep[c.f.][]{chung2010,lecao2011}.

\setcounter{section}{1}

\section{Compression and selection in the GLM framework}
\label{sec:method}

We first define the sparse PLS and introduce a new formulation of the associated optimization problem, based on proximal operator. Contrary to existing approaches, this formulation provides a simple resolution of the covariance maximization problem associated to sparse PLS. Then, we propose an adaptive version of the sparse PLS selection step. Eventually, we will develop our approach to combine sparse PLS and logistic regression.

\subsection{Proximal sparse PLS}

Let $(\mb x_{i}, \xi_i)_{i=1}^n$ be a $n$-sample, with $\mbg\xi = \tr{(\xi_1,\dots,\xi_n)}\in\RR^n$ a continuous response and $\mb x_{i}\in\RR^p$ a set of $p$ covariates, gathered in the matrix $\mb X_{n\times p}=\tr{[\tr{\mb x_{1}},\hdots,\tr{\mb x_{p}}]}$. The PLS solves a linear regression problem. We consider centered data  $\mbg\xi_c$ and $\mb X_c$ to neglect the intercept and the model $\mbg\xi_c = \mb X_c \mbg\beta_{\setminus 0} + \mbg\epsilon$, with the coefficients $\mbg\beta_{\setminus 0}\in\RR^p$. The metric in the observation space $\RR^n$ is weighted by the matrix $\mb V_{n\times n}$.

In the univariate response case, the PLS \citep{boulesteix2007} consists in constructing $K$ components $\mb t_{k}\in\RR^n$ that explain the response, and defined as linear combinations of predictors, i.e.  $\mb t_k = \mb X \mb w_{k}$ with weight vectors $\mb w_{k} \in\RR^p$  ($k=1,\hdots,K$). These weights $\mb w_{k}$ are defined to maximize the empirical covariance of the corresponding components $\mb t_k$ with the response $\mbg\xi_c$. Other PLS algorithms consider the maximization of the squared covariance, however both definitions are equivalent in the univariate response case \citep{dejong1993, boulesteix2007}. To exclude the inherent noise induced by non pertinent covariates in the model, the sparse PLS \citep{lecao2008,chun2010} introduces a variable selection step into the PLS framework. It constructs ``sparse'' weight vectors, whose coordinates are required to be null for covariates that are irrelevant to explain the response. Following the Lasso principle \citep{tibshirani1996}, the shrinkage to zero is achieved with a $\ell_1$ norm penalty in the covariance maximization problem:
\begin{equation}
\hat{\mb w}(\lambda_s) = \underset{\mb w\in\RR^p}\argmin\ \Big\{ -\hat{\text{Cov}}(\mb X_c \mb w, \mbg \xi_c)  +  \lambda_s\,\Vert\mb w\Vert_1\ \Big\},
\label{Eq:SPLSorigine}
\end{equation}
under the constraints $\Vert\mb w\Vert_2 = 1$ and orthogonality between components, with the sparsity parameter $\lambda_s > 0$. The empirical covariance between $\mbg\xi_c$ and $\mb t = \mb X_c\mb w$ is explicitly $\hat{\text{Cov}}(\mb X_c \mb w, \mbg \xi_c) =  \tr{\mb w}\mb c$, where $\mb c = \tr{\mb X_c} \mb V \mbg \xi_c \in\RR^p$ is the empirical covariance $\hat{\text{Cov}}(\mb X_c, \mbg \xi_c)$, depending on the metric weighted by $\mb V$ ($\mb c$ is a vector because the response is univariate).

Different methodologies \citep{lecao2008,chun2010} have been proposed to solve the optimization problem~(\ref{Eq:SPLSorigine}). However, both approaches give an approximate solution. We propose a new approach to exactly solve this problem in the univariate response case. In the standard PLS algorithm, $\mb w$ is proven to be the dominant singular vector of the empirical covariance $\mb c$. In the univariate response case (PLS1 algorithm), $\mb c$ is univariate and $\mb w\propto\mb c$. Thus, we introduce the following equivalent formulation of the penalized problem~(\ref{Eq:SPLSorigine}):
\begin{equation}
\hat{\mb w}(\lambda_s) = \underset{\mb w\in\RR^p}\argmin \ \Big\{ \frac{1}{2} \Vert \mb c - \mb w \Vert_2^{\,2} + \lambda_s\,\Vert \mb w\Vert_1 \Big\},
\label{Eq:SPLS}
\end{equation}
under the constraints $\Vert\mb w\Vert_2 = 1$ and orthogonality between components (the equivalence between (\ref{Eq:SPLSorigine}) and (\ref{Eq:SPLS}) is shown in the Supp. Mat.). We consider a range of values for $\lambda_s$ so that the problem~ (\ref{Eq:SPLS}) admits a solution.

\paragraph{Resolution.} Applying the method of Lagrange multipliers, the problem~(\ref{Eq:SPLS}) becomes ($\mu>0$):
\begin{equation}
\underset{\substack{\mb w\in\RR^p\\ \mu>0}}\argmin \ \Big\{ \frac{1}{2} \Vert \mb c - \mb w \Vert_2^{\,2} + \lambda_s\,\Vert \mb w\Vert_1 + \mu\,\big(\Vert \mb w \Vert_2^{\,2} - 1\big) \Big\}.
\label{Eq:Lagrange}
\end{equation}
The method of Lagrange multipliers was proposed by \cite{witten2009} or \cite{tenenhaus2014} for different decomposition problems. The objective is continuous and convex, thus the strong duality holds and the solutions of primal~(\ref{Eq:SPLS}) and dual~(\ref{Eq:Lagrange}) problems are equivalent. The resolution of the dual problem is based on proximal (or proximity) operators defined as the solution of the following problem \citep{bach2012}:
\begin{equation}
\underset{\mb w\in \RR^p}\argmin \ \Big\{ \frac{1}{2} \Vert \mb c - \mb w\Vert_2^{\,2} + f(\mb w) \Big\},
\label{Eq:Prox}
\end{equation}
for any fixed $\mb c\in\RR^p$, any function $f:\RR^p\to\RR$. It is denoted by $\text{prox}_f(\mb c)$. When $f(\cdot)$ corresponds to the Elastic Net penalty (combination of $\ell_1$ and $\ell_2$ penalty), i.e. $f(\mb w) = \frac{\mu}{2}\,\sum_{j=1}^p \vert w_j\vert^2 + \lambda\,\sum_{j=1}^p \vert w_j\vert$ (with $\lambda>0$ and $\mu>0$), the closed-form solution of problem~(\ref{Eq:Prox}) is explicitely given by the proximal operator $\text{prox}_{\frac{\mu}{2}\,\Vert\cdot\Vert_2^{\,2} + \lambda\,\Vert\cdot\Vert_1}(\mb c)$ whose coordinates are defined by \citep[Theo. 4]{yu2013}:
\begin{equation}
\text{prox}_{\frac{\mu}{2}\,\Vert\cdot\Vert_2^{\,2} + \lambda_s\,\Vert\cdot\Vert_1}(\mb c)= \Big( \frac{1}{1+\mu} \, \text{sgn}(c_j)\,\big(\vert c_j\vert - \lambda\big)_+\Big)_{j=1:p}\, .
\label{Eq:Sol}
\end{equation}
This corresponds to the normalized soft-thresholding operator applied to the covariance vector $\mb c$. When choosing $\mu = \mu^*$ so that $\mb w^* = \text{prox}_{\frac{\mu}{2}\,\Vert\cdot\Vert_2^{\,2} + \lambda\,\Vert\cdot\Vert_1}(\mb c)$ has a unitary norm, the pair $(\mb w^*, \mu^*)$ given by the proximal operator~(\ref{Eq:Sol}) with $\lambda = \lambda_s$ is a candidate point and then a solution (by convexity) for the dual problem~(\ref{Eq:Lagrange}). Hence, the SPLS weights used to construct the SPLS components are given by $\mb w^*\in\RR^p$. This new resolution of the sparse PLS problem is a general result, that also applies in the case of the standard homoskedastic linear model by replacing $\mb V$ by the $n \times n$ identity matrix. This is consistent with the derivation of the sparse PLS by \cite{chun2010}, but provides a more direct resolution framework.  In addition, $\lambda_s$ is renormalized to lie in $[0,1]$ \citep[c.f.][]{chun2010}.

The resolution of the problem~(\ref{Eq:SPLS}) allows to compute $\mb w_1$ and construct the first components $\mb t_1$. At step $k>1$, $\mb w_{k}$ is computed by solving Eq.~\ref{Eq:SPLS}, using a ``deflated'' version of $\mb X_c$ and $\mbg\xi_c$, i.e. the residuals of the respective regression of $\mb X_c$ and $\mbg\xi_c$ onto the previous components $[\mb t_\ell]_{\ell=1}^{k-1}$, guaranteeing the orthogonality between components. The active set of selected variables up to component $K$ is a subset of $\{1,\dots,p\}$, defined as the variables with a non null weights in $[\mb w_{k}]_{k=1}^K$, and denoted by $\mathcal{A}_K = \cup_{k=1}^K \{ j, w_{jk} \ne 0\}$. Eventually, the estimation $\hat{\mbg\beta}^\text{SPLS}_{\setminus 0}$ of $\mbg\beta_{\setminus 0}$ in the model $\mbg \xi_c= \mb X_c \mbg\beta_{\setminus 0} + \mbg\epsilon$ is given by the weighted PLS regression of $\mbg\xi_c$ onto the selected  variables in the active set $\mathcal{A}_K$.  The coefficient $\hat{\beta}^\text{SPLS}_j$ is set to zero if the predictor $j \in\{1,\dots,p\}$ is not in $\mathcal{A}_K$. Indeed, following the definition of the SPLS regression, the sparse structure of the weight vectors $[\mb w_k]_{k=1}^K$ directly induces the sparse structure of $\hat{\mbg\beta}^\text{SPLS}_{\setminus 0}$. The variables selected to construct the new components $[\mb t_k]_{k=1}^K$ are the ones that contribute the most to the response and correspond to those with non-null entries in the true vector $\mbg\beta_{\setminus 0}$.

\subsection{Adaptive sparse PLS}

We also propose to adjust the $\ell_1$ constraint to further penalize the less significant variables, which can lead to a more accurate selection process. Such an approach is inspired by component wise penalization as adaptive Lasso \citep{zou2006a}. We use the weights $\mb w^{\text{PLS}}\in\RR^p$ from classical PLS (without sparsity constraint) to adapt the $\ell_1$ penalty on the weight vector $\mb w^{\text{SPLS}}$. The $\ell_1$ penalty in problem~(\ref{Eq:SPLS}) becomes $\text{Pen}_\text{ada}(\mb w) = \lambda_s \sum_{j=1}^p \gamma^j\, \vert w_j\vert$, with $\gamma^j = 1 / \vert w^{\text{PLS}}_{j} \vert$ to account for the significance of the predictor $j$ (higher weights in absolute values correspond to more important variables). The closed-form solution accounts for the adaptive penalty and remains the soft-thresholding operator applied to $\mb c$ but with parameter $\lambda_s \times \gamma^j$ for $j^{\text{th}}$ predictor (c.f. Supp. Mat.). We called this method adaptive sparse PLS.

\subsection{Ridge-based logistic regression and logistic regression}\label{subsec:logit_spls}

We now present our approach based on sparse PLS for logistic regression.

\paragraph{The Logistic Regression model.} We now consider a $n$-sample $(\mb x_{i}, y_i)_{i=1}^n$ with $y_i$ being a label variable in $\{0,1\}$, gathered in $\mb y = \tr{(y_1,\dots,y_n)}$. We use the Generalized Linear Models (GLM) framework \citep{mccullagh1989} to relate the predictors to the random response variable $Y_i$, using the logistic link function, such that $\text{logit}(\pi_i)=\beta_0 + \tr{\mb x_{i}}\mbg\beta_{\setminus 0},$  with $\pi_i=\EE[Y_i]$, $\text{logit}(x)=\log(x/(1-x))$, and $\mbg \beta=\tr{(\beta_0,\beta_1\hdots,\beta_p)}=\{\beta_0,\mbg\beta_{\setminus 0}\}$. In the sequel, $\mb Z = [\tr{(1,\dots,1)}, \mb X]$. With $\eta_i=\tr{\mb z_{i}}\mbg\beta$, the log-likelihood of the model is defined by $\log\mathcal{L}(\mbg\beta) = \sum_{i=1}^n \left[ y_i \eta_i - \log\{1+\exp(\eta_i)\}\right]$, and the coefficients $\mbg\beta\in\RR^{p+1}$ are estimated by maximum likelihood (MLE). 

\paragraph{The Ridge IRLS algorithm.} Optimization relies on a Newton-Raphson iterative procedure \citep{mccullagh1989} to construct a sequence $(\hat{\mbg\beta}^{(t)})_{t\geq1}$, whose limit $\hat{\mbg\beta}^\infty\in\RR^{p+1}$ (if it exists) is the estimation of $\mbg\beta$. The Iteratively Reweighted Least Squares (IRLS) algorithm \citep{green1984} explicitly defines $(\hat{\mbg\beta}^{(t)})_{t\geq1}$ as the solutions of successive weighted least squares regressions of a pseudo-response $\mbg \xi^{(t)}\in\RR^n$ onto the predictors at each iteration $t$.  The pseudo-response is linearly generated from the predictors based on previous iterations, c.f. Eq~(\ref{Eq:RIRLS}). However, when  $p > n$, the matrix $\mb Z$ is singular, which leads to optimization issues. \cite{lecessie1992} proposed to optimize a Ridge penalized log-likelihood, i.e. $\log\mathcal{L}(\mbg\beta) - (\lambda_R/2)\ \tr{\mbg\beta}\hat{\mb\Sigma}\mbg\beta$, with $\hat{\mb\Sigma}$ the diagonal empirical variance matrix of $\mb Z$ and $\lambda_R>0$ the Ridge parameter. A unique solution of this regularized problem always exists and is computed by the Ridge IRLS (RIRLS) algorithm \citep{eilers2001}, where the weighted regression at each IRLS iteration is replaced by a Ridge weighted regression, hence: 
\begin{equation}
\left\vert
\begin{aligned}
& \hat{\mbg\beta}^{(t+1)} = (\tr{\mb Z} \mb V^{(t)} \mb Z + \lambda_R\, \hat{\mb\Sigma})^{-1} \tr{\mb Z} \mb V^{(t)} \mbg \xi^{(t)},\\
& \mbg \xi^{(t+1)} = \mb Z \hat{\mbg\beta}^{(t)} + \left(\mb V^{(t)}\right)^{-1} \left[\mb y - \mbg \pi^{(t)}\right],\\
\end{aligned}
\right.
\label{Eq:RIRLS}
\end{equation}
with the estimated probabilities $\hat{\mbg \pi}^{(t)} =(\hat{\pi}_i^{(t)})_{i=1}^n$, i.e. $\hat{\pi}_i^{(t)} = \text{logit}^{-1}\big(\tr{\mb z_{i}}\hat{\mbg\beta}^{(t)}\big)$  for each $Y_i$, and $\mb V^{(t)} = \text{diag}\big(\hat{\pi}_i^{(t)}(1-\hat{\pi}_i^{(t)})\big)_{i=1}^n$ is the diagonal empirical variance matrix of $(Y_i)_{i=1}^n$ at step $t$.

Following the definition of $\mbg \xi^{(t)}$, the (R)IRLS algorithm produces a pseudo-response $\mbg \xi^\infty$ as the limit of the sequence $(\mbg \xi^{(t)})_{t\geq1}$, verifying $\mbg \xi^\infty = \mb Z \hat{\mbg\beta}^\infty + \mbg\epsilon$, where $\hat{\mbg\beta}^\infty$ is the solution of the likelihood optimization, and $\mbg\epsilon$ is a noise vector of covariance matrix $(\mb V^\infty)^{-1 }$, with $\mb V^\infty$ the limit of the matrix sequence $(\mb V^{(t)})_{t\geq1}$.

\paragraph{Sparse PLS regression.} The pseudo-response $\mbg \xi^\infty$ produced by Ridge IRLS depends on predictors through a linear model. Following the approach by \cite{fort2005}, we propose to use the sparse PLS regression on $\mbg \xi^\infty$ to process dimension reduction and estimate $\mbg\beta\in\RR^{p+1}$ in the logistic model $\EE[Y_i] = \text{logit}^{-1}(\beta_0 + \tr{\mb x_i}\mbg\beta_{\setminus 0})$. In this case, the $\ell_2$ metric (in the observation space) is weighted by  the empirical inverse covariance matrix $\mb V^\infty$, to account for the heteroskedasticity of noise $\mbg\epsilon$. To neglect the intercept in the SPLS step, we consider the centered version of $\mb X$ and $\mbg\xi^\infty$, regarding the metric weighted by $\mb V^\infty$, denoted by $\mb X_c$ and $\mbg\xi_c^\infty$. The intercept $\beta_0$ will be estimated later.

The estimates $\hat{\mbg\beta}^\text{SPLS}_{\setminus 0}\in\RR^p$ are renormalized to correspond to the non-centered and non-scaled data, i.e. $\hat{\mbg\beta}_{\setminus 0} = \hat{\mbg\Sigma}^{-1/2}\hat{\mbg\beta}^\text{SPLS}_{\setminus 0}$
giving the estimation $\hat{\mbg\beta}_{\setminus 0}$ in the original logistic model . The intercept $\beta_0$ is estimated by $\hat{\beta}_0 = \bar{\xi}^\infty - \tr{\bar{\mb x}}\hat{\mbg\beta}_{\setminus 0}$ where $\bar{\xi}^\infty$ and $\bar{\mb x}$ are respectively the sample average of the pseudo-response and the sample average vector of predictors regarding the metric weighted by $\mb V^\infty$. Our method can be summarized as follow:
\begin{enumerate}
\item $(\mbg \xi^\infty, \mb V^\infty)$ $\gets$ $\text{RIRLS}(\mb X, \mb y, \lambda_R)$
\item Center $\mb X$ and $\mbg\xi^\infty$ regarding the scalar product weighted by $\mb V^\infty$
\item $\left(\hat{\mbg\beta}^\text{SPLS}_{\setminus 0}, \mathcal{A}_K, [\mb t_k]_{k=1}^K\right)$ $\gets$ $\text{SPLS}(\mb X, \mbg \xi^\infty, K, \lambda_s, \mb V^\infty)$
\item Renormalization of $\hat{\mbg\beta} = \{\hat{\beta}_0, \hat{\mbg\beta}_{\setminus 0}\}$
\end{enumerate}
The label $\hat{y}_{\text{new}}$ of new observations $\mb x_{\text{new}}\in\RR^p$ (non-centered and non-scaled) is predicted through the logit function thanks to the estimates $\hat{\mbg\beta} = \{\hat{\beta}_0, \hat{\mbg\beta}_{\setminus 0}\}$. Note that $\mb x_{\text{new}}$ does not need to be centered nor scaled thanks to the intercept parameter $\hat{\beta}_0$ and to the renormalization of the coefficient estimates in the algorithm.

Our method estimates predictor coefficients $\mbg\beta$ in the logistic model by sparse PLS regression of a pseudo-response, considered as continuous and therefore in accordance with the conceptual framework of PLS, while completing compression and variable selection simultaneously. An additional interest is that the iterative optimization in the RIRLS algorithm does not depend on the number of components $K$ nor on the sparsity parameter $\lambda_s$. Consequently, the convergence of our method is robust to the choice of $K$ and $\lambda_s$ by definition, contrary to other approaches for logistic regression based on sparse PLS (c.f. Supp. Mat. section~\ref{supp:sec:comparison}). Our approach will be called logit-SPLS in the following while the method by \cite{fort2005} will be called logit-PLS.

\subsection{Tuning sparsity by stability selection}\label{subsec:stab_sel}
Our logit-SPLS approach depends on three hyper-parameters: the sparsity parameter $\lambda_s$, the Ridge parameter $\lambda_{R}$ and   the number of components $K$. We first propose to tune all the parameters by 10-fold cross-validation (to reduce the sampling dependence). Details about the choice of the grid of candidates values for $(\lambda_s, \lambda_R, K)$ are given in Supp. Mat. (c.f. sections~\ref{supp:subsec:simu} and \ref{supp:subsec:comp_data1}).

In addition, we propose to adapt the stability selection method developed by \cite{meinshausen2010}, to the sparse PLS framework. The interest of this approach is to avoid choosing a value for the sparsity parameter $\lambda_s$ to find the degree of the sparsity in the model, i.e. to select the relevant predictors.
In this framework, the grid of all parameter candidate values for $(\lambda_s, \lambda_R, K)$ is denoted by $\Lambda$. The principle consists in fitting the model for all points $\mbg\ell \in \Lambda$, then estimating the probability $p_j^{\mbg\ell}$ for each covariate $j$ to be selected over 100 resamplings of size $n/2$ depending on ${\mbg\ell}$, i.e. the probability for predictor $j$ to be in the set $\hat{S}_{\mbg\ell} = \{ j,\, \hat{\beta}_j({\mbg\ell}) \ne 0\}$, where $\hat{\mbg\beta}({\mbg\ell})\in\RR^p$ are the corresponding estimated coefficients. Finally, the procedure retains the predictors that are in the set $\hat{S}_\text{stable}$ of stable selected variables, defined as $\{ j, \, \text{max}_{{\mbg\ell}\in\Lambda}\{ p_j^{\mbg\ell}\} \geq \pi_\text{thr}$\}, where $\pi_\text{thr}$ is a threshold value. This means that predictors with high selection probability are kept and predictors with low selection probability are discarded.

The average number of selected variables over the entire grid $\Lambda$, is denoted by $q_\Lambda$, and defined as $q_\Lambda = \EE[\#\{\cup_{\lambda\in\Lambda} \hat{S}_\lambda\}]$. \citet[Theo. 1]{meinshausen2010} provided a bound on the expected number of wrongly stable selected variables (equivalent to false positives) in $\hat{S}_\text{stable}$, depending on the threshold $\pi_\text{thr}$, the expectation $q_\Lambda$ and the number $p$ of covariates:
\begin{equation}
\EE[\text{FP}] \leq \frac{1}{2\pi_\text{thr} -1}\frac{q_\Lambda^2}{p}
\label{Eq:stab_sel}
\end{equation}
where $\text{FP}$ is the number of false positives i.e. $\text{FP} = \#\{ S_0^c \cap \hat{S}_\text{stable}\}$ and $S_0$ the unknown set of true relevant variables. This results is derived under some reasonable conditions that are discussed in Supp. Mat. (section~\ref{supp:sec:cond_stab_sel}). Following the recommendation of \citet[p. 424]{meinshausen2010}, we use Eq.~\ref{Eq:stab_sel} to determine the range of the parameter grid $\Lambda$ to avoid too many false positives (corresponding to a weak $\ell_1$ penalization). Indeed, since the number of false positives is controlled by $q_\Lambda$, we automatically exclude candidate points $\mbg\ell = (\lambda_s, \lambda_R, K)$ corresponding to small $\lambda_s$ (near 0)  for which there is no selection and for which all variables contribute to the mode, so that we can control $q_\Lambda$. Without removing these points, $q_\Lambda$ and the number of false positives are too high. For instance, when the threshold probability $\pi_\text{thr}$ is set to 0.9, $\Lambda$ is defined as a subset of the parameter grid so that $q_\Lambda = \sqrt{0.8\,p\,\rho_\text{error}}$. In practice, $q_\Lambda$ is unknown but can be estimated by the empirical average number of selected variables over all $\mbg\ell\in\Lambda$. In this context, the expected number of false positives will be lower than $\rho_\text{error}$ (in practice, we set $\rho_\text{error}=10$). Details about the candidate values for $(\lambda_s, \lambda_R, K)$ are given in Supp. Mat. (section \ref{supp:subsec:stab_sel_data1}).

A clear interest here is that we do not have to choose a specific value for the hyper-parameters, instead we retain the variables that are selected by most of the models when exploring the grid of candidate values for hyper-parameters (including $K$).

\section{Simulation study}
\label{sec:simu}

We assess the performance of our approach for prediction, compression and variable selection compared to state-of-the-art methods that were previously introduced. We also use a ``baseline'' method, called GLMNET \citep{friedman2010}, that performs variable selection, by solving the GLM likelihood maximization with $\ell_1$ norm penalty for selection and $\ell_2$ norm penalty for regularization, also known as the Elastic Net approach \citep{zou2005}. We compare different approaches based on (sparse) PLS for classification (c.f. Tab.~\ref{tab:algo} and Supp. Mat. section \ref{supp:sec:comparison} and \ref{supp:sec:imp} for details).

\begin{table*}[!t]
\processtable{The different algorithms to process dimension  reduction by (sparse) PLS in the framework of the logistic regression.\label{tab:algo}}{
\begin{tabular}{>{\raggedright\arraybackslash}p{16mm}@{\hskip 1mm}>{\raggedright\arraybackslash}p{60mm}@{\hskip 4mm}c@{\hskip 4mm}>{\raggedright\arraybackslash}p{50mm}}
\toprule
Method & Algorithm & Sparse? & Reference \\
\midrule
GPLS & \multirow{2}{70mm}[-0.2em]{(S)PLS inside the IRLS algorithm} & $\times$ & \cite{ding2005} \\
SGPLS & & $\checkmark$ & \cite{chung2010} \\
\midrule
PLS-log & \multirow{2}{70mm}[-0.4em]{(S)PLS before logistic regression}  & $\times$ & \cite{wang1999}, \cite{nguyen2002}\\
SPLS-log & & $\checkmark$ & \cite{chung2010} \\
\midrule
logit-PLS & \multirow{2}{70mm}[-0.3em]{(S)PLS on the pseudo-response after the RIRLS algorithm} & $\times$ & \cite{fort2005} \\
logit-SPLS & & $\checkmark$ & \textbf{Our algorithm} \\
\botrule
\end{tabular}}{}
\end{table*}

\paragraph{Simulation design.} Our simulated data are constructed to assess the interest of compression and variable selection for prediction performance. The simulations are inspired from \cite{zou2006}, \cite{shen2008} or \cite{chung2010}. The purpose is to control the redundancy within predictors, and the relevance of each predictor to explain the response. We consider a predictor matrix $\mb X$ of dimension $n\times p$, with $n=100$ fixed, and $p=100, 500, 1000, 2000$, so that we examine different high dimensional models. The true vector coefficients $\mbg\beta^*$ is generated to be sparse,  the sparsity structure is thus known. Hence, it is possible to assess whether a method selects the relevant predictors or not. The response variable $Y_i$ for observation $i$ is a Bernoulli variable, with parameter $\pi_i^* = \text{logit}^{-1}(\tr{\mb x_{i}}\mbg\beta^*)$. The pattern of data simulation and the tuning of hyper-parameters are detailed in Supp. Mat. (section \ref{supp:sec:comp_simu}). Regarding other methods, we use the range of parameters recommended by their respective authors and the cross-validation procedures supplied in the corresponding packages.

\paragraph{Ridge penalty ensures convergence.} Convergence is crucial when combining PLS and IRLS algorithm as pointed by \cite{fort2005}. With the analysis of high dimensional data and the use of selection in the estimating process, it becomes even more essential to ensure the convergence of the optimization algorithms, otherwise the output estimates may not be relevant. Our simulations show that the Ridge regularization systematically ensures the convergence of the IRLS algorithm in our method (logit-SPLS), for any configuration of simulation: $p = n$, $p > n$, high or low sparsity, high or low redundancy (see Tabs.~\ref{supp:tab:simu:conv_cv} and \ref{supp:tab:simu:conv} in Supp. Mat.). On the contrary, approaches that use (sparse) PLS before or within the IRLS algorithm (resp. SPLS-log and (S)GPLS) encounter severe convergence issues.

Whereas the SPLS-log or (S)GPLS approaches were designed to overcome convergence issues, it appears that they do not, which questions the reliability of the results supplied by these methods. Then, it confirms the interest of the Ridge regularization to ensure the convergence of the IRLS algorithm. Moreover, this convergence seems to be fast (around 15 iterations even when $p=2000$), which depicts an interesting outcome for computational time. For instance, the tuning of three parameters in the logit-SPLS approach is less costly thanks to the fast convergence of the algorithm. Although both SGPLS and SPLS-log methods are based on two parameters, they iterates further (until the limit set by the user) which is less computationally efficient, especially with high dimensional data. On this matter, details regarding computation times are given in Supp. Mat. (section~\ref{supp:add_res_simu}).

\paragraph{Adaptive selection improves cross-validation stability.} A cross-validation procedure would be expected to be stable under multiple runs, i.e. the chosen values must not be variable when running the procedure many times on the same sample. Otherwise, selection and prediction become uncertain and not suitable for experiment reproducibility. We quantified the standard deviation of the sparse parameter $\lambda_s$ chosen by cross-validation for the three sparse PLS methods (SGPLS, SPLS-log and our logit-SPLS) when repeating the procedure on the same samples. The standard deviation (all three methods consider the same range of values for $\lambda_s$) is smaller for our approach (c.f. Tab.~\ref{tab:simu:stab}) than for other methods. Thus, the cross-validation procedure in our adaptive method is more stable than other SPLS approaches. A similar comment can be made regarding the choice of the number $K$ of components (c.f. Fig.~\ref{supp:fig:simu:cv_K} in Supp. Mat.). This behavior can be linked to the convergence of the different approaches. The methods with convergence issues (SGPLS and SPLS-log) present a higher cross-validation instability, whereas our method (logit-SPLS) converges efficiently and shows a better cross-validation stability.
Similarly, the variable selection accuracy, defined as the proportion of rightly selected and rightly non selected variables \citep{chong2005}, is also influenced by the cross-validation stability and the convergence of the method. Indeed, the standard deviation of the selection accuracy (computed across multiple runs) is higher for the less stable and less convergent methods (SGPLS and SPLS-log) compared to our logit-SPLS approach (c.f. Tab.~\ref{tab:simu:stab}).

\setlength{\tabcolsep}{14pt}
\begin{table}[!t]
\processtable{Comparing computational stability between sparse PLS approaches. $\hat{\sigma}\big(\hat{\lambda}_s\big)$ stands for the estimated standard deviation of the tuned hyper-parameter $\lambda_s$ (over repetitions on the same simulated data set), which measures the stability of the hyper-parameter tuning by cross-validation. $\hat{\sigma}\big(\text{acc.}\big)$ stands for the estimated standard deviation of the accuracy in variable selection, which measures the stability of the selection steps. The results are presented for different model dimensions ($p$). \label{tab:simu:stab}} {\begin{tabular}{@{}lcccc@{}}
\toprule
\multirow{2}{*}{Method} & \multicolumn{2}{c}{$p=100$} &  \multicolumn{2}{c}{$p=2000$} \\
 & $\hat{\sigma}\big(\hat{\lambda}_s\big)$ & $\hat{\sigma}\big(\text{acc.}\big)$ & $\hat{\sigma}\big(\hat{\lambda}_s\big)$ & $\hat{\sigma}\big(\text{acc.}\big)$ \\
\midrule
\textbf{logit-spls} & $\boldsymbol{0.09}$ & $\boldsymbol{0.11}$ & $\boldsymbol{0.11}$ & $\boldsymbol{0.09}$\\
sgpls & 0.17 & 0.14 & 0.15 & 0.12 \\
spls-log & 0.23 & 0.12 & 0.21 & 0.17 \\
\botrule
\end{tabular}}{}
\end{table}
\setlength{\tabcolsep}{3pt}

\paragraph{Compression and selection increase prediction accuracy.} We now assess the importance of compression and variable selection for prediction performance. We consider the prediction accuracy, evaluated through the prediction error rate. A first interesting point is that the prediction performance of compression methods is improved by the addition of a selection step: logit-SPLS, SGPLS and SPLS-DA perform better than logit-PLS, GPLS and PLS-DA respectively (c.f. Tab.~\ref{tab:simu:pred_sel}). In addition, sparse PLS approaches also present a lower classification error rate than the GLMNET method that performs variable selection only. These two points support our claim that in any case compression and selection should be both considered for prediction. Similar results are observed for other configurations of simulated data (c.f. Supp. Mat. section~\ref{supp:add_res_simu}). All different SPLS-based approaches show similar prediction performance, even methods that are not converging (SPLS-log or SGPLS) compared to our adaptive approach logit-SPLS. Thus, checking prediction accuracy only may not be a sufficient criterion to assess the relevance of a method. The GPLS method is a good example of non-convergent method (c.f. Tab.~\ref{tab:simu:pred_sel} and Tab.~\ref{supp:tab:simu:conv} in Supp. Mat.) that presents high variability and poor performance regarding prediction.

Actually, the combination of Ridge IRLS and sparse PLS in our method ensures convergence and provides good prediction performance (prediction error rate at 10\% on average) even in the most difficult configurations $n=100$ and $p=2000$, which makes it an appropriate framework for classification.

\setlength{\tabcolsep}{8.5pt}
\begin{table}[!t]
\processtable{Prediction error and selection sensitivity/specificity (if relevant) when $p=2000$, for non-sparse or sparse approaches (delimited by the line). Results for other values of $p$ are joined in Supp. Mat. (section~\ref{supp:add_res_simu}). \label{tab:simu:pred_sel}} {\begin{tabular}{@{}lcccc@{}}
\toprule
\multirow{2}{*}{Method} & Prediction & Selection & Selection & Selection \\
& error & sensitivity & specificity & accuracy \\
\midrule
gpls & $0.49 \pm 0.31$ & / & / & / \\
pls-da & $0.20 \pm 0.07$ & / & / & / \\
logit-pls & $0.17 \pm 0.07$ & / & / & / \\
\midrule
glmnet & $0.16 \pm 0.07$ & $0.27$ & $0.98$ & 0.74 \\
\textbf{logit-spls} & $\mbg{0.11 \pm 0.06}$ & $\mbg{0.63}$ & $\mbg{0.86}$ & $\mbg{0.79}$ \\
sgpls & $0.11 \pm 0.05$ & $0.80$ & $0.75$ & $0.81$ \\
spls-da & $0.12 \pm 0.06$ & $0.82$ & $0.74$ & $0.81$ \\
spls-log & $0.12 \pm 0.05$ & $0.83$ & $0.75$ & $0.81$\\
\botrule
\end{tabular}}{}
\end{table}
\setlength{\tabcolsep}{3pt}

\paragraph{Compression increases selection accuracy.} A sparse model will be useful if characterized by good prediction performances but also if the selected covariates are the genuine important predictors that explain the response. To assess the selection accuracy, we compare the selected predictors returned by each sparse method to the set of relevant ones used to construct the response, i.e. with a non zero coefficient $\beta^*_j$ in our simulation model.  We consider sensitivity and specificity \citep{chong2005}, respectively the proportion of true positive and true negative regarding the selected variables.

A first striking point is that, in our simulations (see Tab.~\ref{tab:simu:pred_sel} and Tabs.~\ref{supp:tab:simu:pred_sel1},~\ref{supp:tab:simu:pred_sel2},~\ref{supp:tab:simu:pred_sel3} in Supp. Mat.), the baseline GLMNET presents a very low sensitivity and a very high specificity (low true positive and low false positive rates), meaning that it selects a small number of predictors (that are relevant), which leads to a lower accuracy compared to SPLS-based approaches. Thus, using approaches that combine compression and variable selection such as sparse PLS has a true impact on selection accuracy, compared to ``selection-only'' approach such as GLMNET.

Then, we focus on the comparison of the different sparse PLS approaches. On the one hand, our method logit-SPLS selects less irrelevant predictors since the false positive rate is lower (higher specificity), compared to other SPLS approaches. On the other hand, SGPLS, SPLS-log and SPLS-DA select more true positives (higher sensitivity). Since all methods achieve a similar level of accuracy, this result clearly illustrates a difference of strategy regarding variable selection. The balance between sensitivity and specificity indicates that our method logit-SPLS selects predictors which are more likely to be relevant, discarding most of the non-pertinent predictors, while other approaches tend to select more predictors with higher false positive rate. With high dimensional data set (large $p$), we are generally interested in highly sparse model, thus it is an advantage to have a sharper control on the false positive rate, as in our method. In addition, the relative good sensitivity of other sparse PLS approaches (SGPLS and SPLS-log) is also balanced by a selection process that is less stable than ours, as the standard deviation of the accuracy is higher over simulations (as previously mentioned, see Tab.~\ref{tab:simu:stab}).

\section{Classification of breast tumors using adaptive sparse PLS for logistic regression}
\label{sec:data}

We consider a publicly available data set on breast cancer \citep{guedj2012} containing the level expression of 54613 genes for 294 patients affected by breast cancer. We focus on the relapse after 5 years, considering a $\{0,1\}$ valued response, if the relapse occurred or not. There were 214 patients without relapse and 80 with a relapse. We reduce the number of genes by considering the top 5000 most differentially expressed genes, by using a standard t-test with a Benjamini-Hochberg correction. Computation details (resamplings, cross-validation, stability selection, training and test set definition) are joined in Supp. Mat. (c.f. section~\ref{supp:sec:comp_data1}).

\paragraph{Convergence and stability with Ridge IRLS and adaptive sparse PLS.} The Ridge IRLS algorithm confirms its usual convergence (see Tab.~\ref{tab:data:res}). Other approaches based on SPLS (SGPLS and SPLS-log) again encounter severe issues and almost never converge. Following a similar pattern, our adaptive selection is far more stable under the tuning of the sparsity parameter $\lambda_s$ by cross-validation than any other approach using sparse PLS (Tab.\ref{tab:data:res}), as the precision on this hyper-parameter value is the highest for our method, illustrating less variability in the tuning over repetitions.

\setlength{\tabcolsep}{14pt}
\begin{table}[!t]
\processtable{Averaged prediction error, convergence percentage over 100 resamplings and standard deviation of cross-validated $\lambda_s$.  \label{tab:data:res}} {\begin{tabular}{@{}lccc@{}}
\toprule Method & Prediction error & Conv. perc. & s.d. $\lambda_s$ \\
\midrule
glmnet & $0.27 \pm 0.04$ & / & / \\
logit-pls & $0.26 \pm 0.05$ & $100\%$ & / \\
logit-spls & $0.23\pm 0.06$ & $100\%$ & $0.15$ \\
\textbf{logit-spls-ad} & $\boldsymbol{0.19\pm 0.04}$ & $\boldsymbol{100\%}$ & $\boldsymbol{0.15}$ \\
sgpls & $0.5 \pm 0.21$ & $5\%$ & $0.18$ \\
spls-log & $0.18\pm0.04$ & $1\%$ & $0.19$ \\
\botrule
\end{tabular}}{}
\end{table}
\setlength{\tabcolsep}{3pt}

\paragraph{Interest of adaptive selection for prediction and selection.} Regarding prediction performance, the adaptive version of our algorithm logit-SPLS gives better results (c.f. Tab.~\ref{tab:data:res}) which highlights the interest of adaptive selection. It can also be noted that our approach performs better on prediction than both logit-PLS (compression only) and GLMNET (selection only), which again supports the interest of using both compression and variable selection. The SGPLS method does not confirm its performance on our simulatiosn with poor and highly variable results, illustrating the potential lack of stability of non-convergent method. Only the SPLS-log method achieves a classification that is as good as our adaptive method. However this point will be counterbalanced by its assessment over the other criteria in the following.

\begin{figure}[!tpb]
\includegraphics[width=0.99\linewidth]{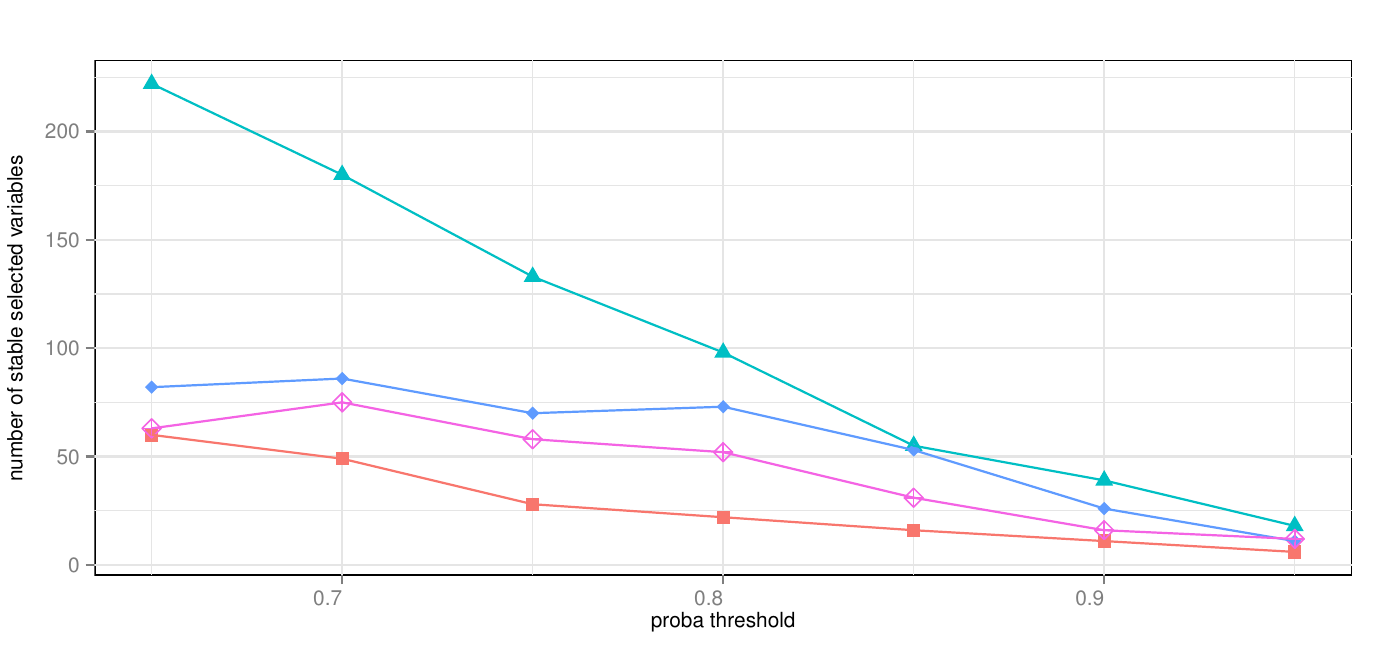}
\caption{Number of variables in the set of stable selected variables versus the threshold $\pi_{\text{thr}}$, when forcing the average number of false positives to be smaller than $\rho_\text{error}=10$. Methods: glmnet ($-${\tiny$\blacksquare$}$-$), logit-spls-adapt ($-$$\blacktriangle$$-$), sgpls ($-$$\diam$$-$), spls-log ($-$$\btimes$$-$).  Note: here, all hyper-parameters (including $K$) vary across the grid of candidate values $\Lambda$ (c.f. Supp. Mat. section \ref{supp:subsec:stab_sel_data1}).}
\label{fig:data:sel}
\end{figure}

Regarding variable selection, the stability selection analysis (see Fig.~\ref{fig:data:sel}) shows that, when the number of false positives is bounded (on average), our approach logit-SPLS selects more genes than any other approach (SGPLS, SPLS-log and GLMNET). Hence, we discover more true positives (because the number of false positives is bounded), unraveling more relevant genes than other approaches. This again illustrates the good performance of our method for selection. More generally, approaches that use sparse PLS, i.e. performing selection and compression, select more variables than GLMNET with the same false positive rate, thus retrieving more true positives than GLMNET which performs only selection. This again supports our previously developed idea that compression and selection are both very suitable for high dimensional data analysis. We recall that the curves in Fig.~\ref{fig:data:sel} correspond to the number of variables that are selected by most models when exploring the grid of candidate values for hyper-parameters (including $K$). Additional results regarding the overlap between the genes selected by the different methods and the list of selected genes with their score (i.e. the maximum estimated probability of selection) are given in Supp. Mat. (section~\ref{supp:subsec:stab_sel_data1}).

\begin{figure}[!tpb]
\centering
\includegraphics[width=0.99\linewidth]{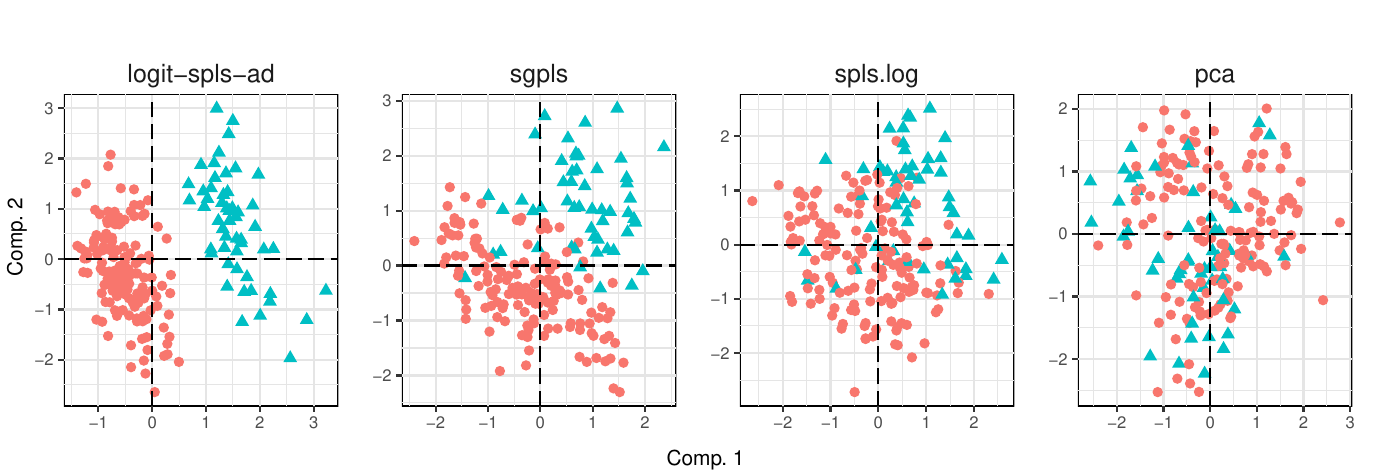}
\caption{\small Observations scores on the first two components for the different methods. The points are shaped according to the value of the response: 0 ({\large$\bullet$}) and 1 ($\blacktriangle$). The scores are normalized for comparison.}
\label{fig:data:comp}
\end{figure}

\paragraph{Efficient compression to discriminate the response.} To assess the interest of our approach for data visualization, we represent the score of the observations on the first two components, i.e. the point cloud $(t_{i1},t_{i2})_{i=1}^n$. The points are colored according to their $Y$-labels. An efficient compression technique would separate the $Y$-classes with fewer components. We fit the different compression-based approach (when the number of components is set to $K=2$). We use PCA as a reference for compression and data visualization, based on unsupervised learning contrary to other compared approaches. Fig.~\ref{fig:data:comp} represents the first two components computed by logit-PLS, logit-SPLS, SGPLS, SPLS-log and PCA. It appears that the first two components from our logit-SPLS are sufficient to easily separate the two $Y$-classes. On the contrary, other sparse PLS approaches do not achieve a similar efficiency in the compression process. Thus, our method turns out to be very efficient for data visualization, especially compared to principal component analysis.

\section{Characterization of T lymphocyte types based on single-cell data}

We generalized our approach to the multicategorial case and developed a new method, called multinomial-SPLS (or MSPLS), that was applied to the prediction of cell types using single cell expression data \citep{stegle2015, gawad2016}. Our approach (detailed in Supp. Mat., section~\ref{supp:sec:multinom_spls}) is based on a direct extension of the logistic model. It is specifically a ``one-class vs a reference'' type of multi-classification, in which the membership probabilities of each class (except the reference) are estimated based on linear combinations of the predictors. The membership probability of the reference class is then deduced from the rest. The resolution is derived from our logit-SPLS method. One interest is that our multi-group classification approach uses a univariate response sparse PLS algorithm (that admits a closed-form solution, c.f. section~\ref{sec:method}), contrary to sparse multigroup PLS-DA for instance (c.f. Supp. Mat. section~\ref{supp:sec:comparison}).

Understanding the mechanisms of an adaptive immune response is of great interest for the creation of new vaccines. This response is made possible thanks to antigen-specific ``effector'' T cells capable of recognizing and killing infected cells, and to the long-lasting ``memory'' T cells that will constitute a repertoire for later secondary immune responses. These two types of T cells have then been described as 4 sub-groups: CM, TSCM (``Memory''), TEMRA, EM, (``Effector Memory''). Generally speaking, CM and TSCM can be considered as ``Memory'' cells and TEMRA and EM can be considered as ``Effectors'' as CM/TSCM and EM/TEMRA share significant functional overlap with each other \citep{WFH15,GLJ11}. Understanding the transcriptomic diversity of T cells constitutes a new challenge to better characterize the short and long-term vaccinal responses, as T cells are increasingly recognized as being highly heterogeneous populations \citep{NSB12}. However, these investigations have been limited by current practices that consist in defining those 4 cell types based by drawing non-overlapping gates on the 2D-space defined two surface markers only: CCR7 and CD45RA \citep{SLF99}. Consequently this rule leads to the selection of a fraction of cells that only correspond to cells with the most extreme values of markers, which ignores the complexity of a T cell population sampled from real blood.

We developed a SPLS-based multi-categorical classification to better characterize the transcriptomic diversity that supports the 4 different cell types of T cells. This approach aims at classifying more cells, and at inferring the type of the non-identified cells. To do so, we considered the measurements of 11 surface markers (CCR7, CD45RA, CD27, IL7R, FAS, CD49F, PD1, CD57, CD3E, CD8A), along with the expression of the corresponding genes. All these measurements were available on the single-cell basis. We will show that even in this low dimensional case, the use of variable selection will help to improve the accuracy of the results. In the following, hyper-paramaters (including $K$) were tuned by cross-validation. Details about the candidate values for $(\lambda_s, \lambda_R, K)$ are given in Supp. Mat. (section \ref{supp:subsec:comp_data2}).

We developed the following two-step analysis. We started by considering the measures of the 11 surface markers and the expression of the 11 associated genes. The multinomial-SPLS was trained on a subset of cells that were tagged manually, and used to predict the types of the unknown cells (136 annotated over 943 cells). On this training set of 136 cells, including 44 CM and 28 TSCM cells (i.e. 72 ``Memory'' cells), 30 EM and 34 TEMRA cells (i.e. 64 ``Effector'' cells), a 5-fold cross-validation procedure (with 50 repetitions) is used to tune the hyper-parameters. The cross-validation prediction error over the resamplings was $\sim 6\%$.Fig.~\ref{supp:fig:data:cell_type2} in Supp. Mat. shows that the cells in the training set are well discriminated in this first step. In addition, our SPLS procedure selected the proteins  CCR7 and CD45RA in 100\% of the runs, which is coherent with the manual annotation of the cells based on these two markers.

In a second step we wanted to enrich the set of genes that discriminate cell types. To proceed we considered the expression of all genes of  these predicted cell types, and performed a differential analysis from  which we retained 61 differentially expressed genes (corresponding to a 5\% FDR). By considering these 61 genes added to the first 22 markers considered for the first prediction step, we performed the MSPLS-based prediction on the complete data set annotated by our first prediction. Our method selected 8 new biologically relevant genes (more details in Supp. Mat. section \ref{supp:subsec:add_data2}) with a cross-validation prediction error rate over re-samplings (again 5-fold cross-validation) of $\sim 16\%$ (on the whole data set, not only considering the most extreme phenotypes). The main interests of this two-step procedure were to be computationally efficient and to narrow the list of potential genes of interest, which was conclusive since this second prediction greatly improved the biological relevance of the predicted cell types by accounting for more information than the one contained in the classical markers like CCR7 and provided us with new insight to better understand the T cells immune response.

\begin{figure}[!tpb]
\centering
\includegraphics[width=0.99\linewidth]{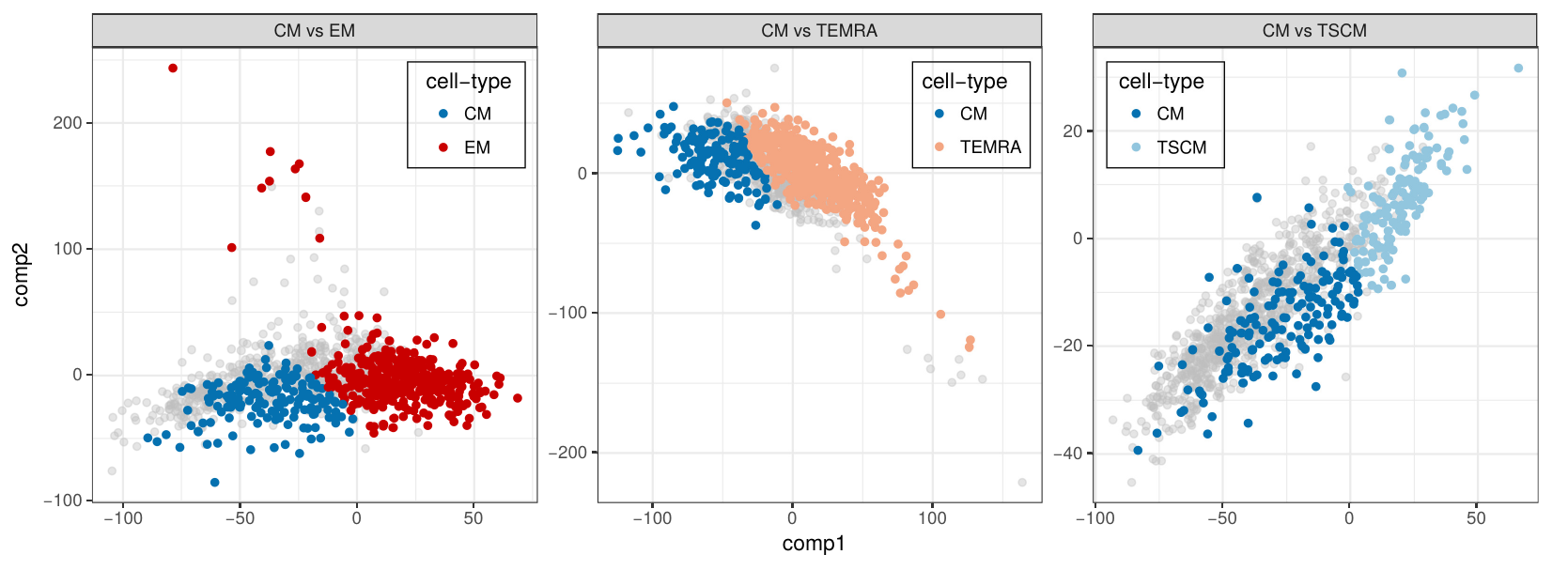}
\caption{\small Cell scores on the first two PLS components in the latent space that discriminate between the reference class (``CM'') and each other class separately (``EM'', ``TEMRA'' and ``TSCM'' respectively, from left to right). T cells are identified by their predicted types after the second prediction step.}
\label{fig:data:cell_type}
\end{figure}

Fig.~\ref{fig:data:cell_type} illustrates the representation of the cells in the latent dimensional space computed by the multinomial PLS in the second step of prediction. The reference class is ``CM''. The SPLS computes latent directions discriminating each other class (``EM'', ``TEMRA'' and ``TSCM'' respectively) versus the reference class (c.f. Supp Mat.). The cells are represented on the first two components for the three different pairs: ``CM versus EM'', ``CM versus TEMRA'' and ``CM versus TSCM''. The latent components clearly discriminate the group of cells in the three different cases, which confirms the result of the second prediction based both on markers and differentially expressed genes. The different groups are clearly identified but there is no gap between them, contrary to the representation of the cells in the training set for the the first prediction (c.f. Supp. Mat.). This indicates that the multinomial-SPLS was able to predict the type of the lost common cells based on the most extreme phenotypes.

This application highlights the interest of dimension reduction by compression and variable selection, even when dealing with low dimensional data. It can also be noted that, even when using sparse approaches, a step of pre-selection is always useful, especially in the analysis of single-cell expression data, which are very noisy compared to standard RNA-seq data, because of the important inter-cellular diversity.

\section{Conclusion}
\label{sec:conc}

We have introduced a new formulation of sparse PLS and proposed an adaptive version of our algorithm to improve the selection process. Using proximal operators, we provide an explicit resolution framework with a closed-form solution based on soft-thresholding operators.

In addition, we developed a method that performs compression and variable selection suitable for classification. It combines Ridge regularized Iterative Least Square algorithm and sparse PLS in the logistic regression context.  It is particularly appropriate for the case of high dimensional data, which appears to be a crucial issue nowadays, for instance in genomics. Our main consideration was to ensure the convergence of IRLS algorithm, which is a critical point in logistic regression. Another concern was to properly incorporate into the GLM framework a dimension reduction approach such as sparse PLS.

Ridge regularization ensures the convergence of the IRLS algorithm, which is confirmed in our simulations and tests on experimental data sets. Applying adaptive sparse PLS as a second step on the pseudo-response produced by IRLS respects the definition of PLS regression for continuous response. Moreover, the combination of compression and variable selection increases the prediction performance and selection accuracy of our method, which turns out to be more efficient than state-of-the-art approaches that do not use both dimension reduction techniques. Such a combination also improves the compression process, illustrated by the efficiency of our method for data visualization compared to standard supervised or unsupervised approaches. Furthermore it appears that previous procedures using sparse PLS with logistic regression encounter convergence issues linked to a lack of stability inthe  cross-validation parameter tuning process, highlighting the crucial importance of convergence when dealing with iterative algorithms.

It can be noted that our approach can be used to include additional covariates in the model. For example, we used a combination of surface marker levels and gene expression levels in the single cell data analysis. On this matter, an interesting research direction would be to work on a Least Square-Partial Least Square (LS-PLS) approach, in which some part of the predictors are compressed into PLS components and some others are not. There have been recent advances regarding LS-PLS for logistic regression \citep[see][]{bazzoli2016}. However, to our knowledge, there is no work on a potential LS-SPLS method, even in the regression case.

In addition, an interesting extension of our work would be to investigate theoretical properties of the sparse PLS regression (especially regarding its consistency or any oracle properties). Deriving such properties would be an opportunity to assess the underlying statistical properties of our method and remains an open question.\vspace*{-10pt}

\section*{Funding}

This work was supported by the french National Resarch Agency (ANR) as part of the ``Algorithmics, Bioinformatics and Statistics for Next Generation Sequencing data analysis'' (ABS4NGS) ANR project [grant number ANR-11-BINF-0001-06] and as part of the ``MACARON'' ANR project [grant number ANR-14-CE23-0003]. It was performed using the computing facilities of the computing center LBBE/PRABI.\vspace*{-12pt}

\bibliographystyle{natbib}
\bibliography{article_logit_spls_2017}

\newpage


\setcounter{section}{0}
\setcounter{figure}{0}
\setcounter{table}{0}
\setcounter{equation}{0}

\renewcommand{\thetable}{A.\arabic{table}}
\renewcommand{\thefigure}{A.\arabic{figure}}
\renewcommand{\thesection}{A.\arabic{section}}
\renewcommand{\theequation}{A.\arabic{equation}}

\section*{Supplementary Information}

\section{Optimization in sparse PLS}


\subsection{Reformulation of the sparse PLS problem}
As previously introduced, the sparse PLS constructs components as sparse linear combination of the covariates. When considering the first components, i.e. $\mb t_1 = \mb X\mb w_1$, the weight vector $\mb w_1\in\RR^p$ is defined to maximize the empirical covariance between the component and the response, i.e. $\hat{\Cov}(\mb X\mb w, \mbg\xi) \propto \tr{\mb w}\tr{\mb X_c}\mbg\xi_c$ (centered $\mb X$ and $\mbg\xi$) with a penalty on the $\ell_1$-norm of $\mb w_1$ to enforce sparsity in the weights. Thus, the weight vector $\mb w_1$ is computed as the solution of the following optimization problem:
\begin{equation}
\left\{
\begin{aligned}
& \underset{\mb w\in\RR^p}\argmin \ \Big\{  - \tr{\mb w}\tr{\mb X_c}\mbg\xi_c + \lambda_s \sum_j \vert w_j \vert\, \Big\}, \\
& \Vert \mb w \Vert_2 = 1 \ \text{(additional constraint)}\, ,
\end{aligned}\right.
\label{appendix:eq:spls}
\end{equation}
with $\lambda_s>0$. The problem~(\ref{appendix:eq:spls}) is equivalent to the following, when denoting the standard scalar product by $\langle\cdot,\cdot\rangle$:
\[
\left\{
\begin{aligned}
& \underset{\mb w\in\RR^p}\argmin \ \Big\{ - 2 \, \big\langle \mb w\, ,\, \tr{\mb X_c}\mbg\xi_c \big\rangle + \Vert \mb w \Vert_2^{\,2} + 2\lambda_s \sum_j \vert w_j \vert\, \Big\}, \\
& \Vert \mb w \Vert_2 = 1\, ,
\end{aligned}\right.
\]
because the term $\Vert \mb w \Vert_2$ is constant thanks to the additional constraint. This new problem remains equivalent to the following:
\[
\left\{
\begin{aligned}
& \underset{\mb w\in\RR^p}\argmin \ \Big\{ \Vert \tr{\mb X_c}\mbg\xi_c \Vert_2^{\,2} - 2 \, \big\langle \mb w\, ,\, \tr{\mb X_c}\mbg\xi_c \big\rangle + \Vert \mb w \Vert_2^{\,2} + 2\lambda_s \sum_j \vert w_j \vert\, \Big\}, \\
& \Vert \mb w \Vert_2 = 1\, ,
\end{aligned}\right.
\]
since the norm of the empirical covariance $\Vert \tr{\mb X_c}\mbg\xi_c \Vert_2^{\,2}$ is constant. Then, thanks to the Euclidean norm properties, it can be rewritten as:
\begin{equation}
\left\{
\begin{aligned}
& \underset{\mb w\in\RR^p}\argmin \ \Big\{ \frac{1}{2} \Vert \mb c - \mb w \Vert_2^{\,2} + \lambda_s\,\Vert \mb w\Vert_1\, \Big\}, \\
& \Vert \mb w \Vert_2 = 1\, ,
\end{aligned}\right.
\label{appendix:eq:spls2}
\end{equation}
with $\mb c = \tr{\mb X_c}\mbg\xi_c$ and when setting $\lambda = 2\nu>0$. Actually, in the case of a univariate response, the formulation~(\ref{appendix:eq:spls2}) is natural. Indeed, in the standard (non-sparse) PLS, the optimal weight vector $\mb w$ is the normalized dominant singular vector of the covariance matrix $\tr{\mb X}\mbg\xi$. However, when the response is univariate, the matrix $\tr{\mb X}\mbg\xi$ is a vector and the solution for $\mb w$ is the normalized vector $\tr{\mb X}\mbg\xi$ (normalized to 1). This corresponds exactly to the solution of the problem:
\[
\left\{
\begin{aligned}
& \underset{\mb w\in\RR^p}\argmin \ \Vert \mb c - \mb w \Vert_2^{\,2}\, , \\
& \Vert \mb w \Vert_2 = 1\, ,
\end{aligned}\right.
\]
(without the $\ell_1$ penalty).

The solution of the penalized problem~(\ref{appendix:eq:en}) defines the first component ($k=1$) of the sparse PLS. We use deflated predictors and response to construct the following component ($k>1$).

\subsection{Resolution of the sparse PLS problem}

Applying the method of Lagrange multipliers, the problem~(\ref{appendix:eq:spls2}) becomes:
\begin{equation}
\underset{\substack{\mb w\in\RR^p\\ \mu>0}}\argmin \ \Big\{ \frac{1}{2} \Vert \mb c - \mb w \Vert_2^{\,2} + \lambda_s\,\Vert \mb w\Vert_1 + \mu\,\big(\Vert \mb w \Vert_2^{\,2} - 1\big)\Big\},
\label{appendix:eq:en}
\end{equation}
with $\mu >0$. The objective is continuous and convex, thus the strong duality holds and the solutions of primal and dual problems are equivalent.

To solve the problem~\ref{appendix:eq:en}, we use proximity operator (also called proximal operator) defined as the solution of the following problem \citep{bach2012}:
\begin{equation}
\underset{\mb w\in \RR^p}\argmin \ \Big\{ \frac{1}{2} \Vert \mb c - \mb w\Vert_2^{\,2} + f(\mb w) \Big\},
\label{appendix:eq:prox0}
\end{equation}
for any fixed $\mb c\in\RR^p$, any function $f:\RR^p\to\RR$. It is denoted by $\text{prox}_f(\mb c)$. When $f(\cdot)$ corresponds to the Elastic Net penalty (combination of $\ell_1$ and $\ell_2$ penalty), i.e. when considering the problem (with $\lambda>0$ and $\mu>0$):
\begin{equation}
\underset{\mb w\in\RR^p}\argmin \ \Big\{ \frac{1}{2} \, \Vert \mb c - \mb w \Vert_2^{\,2} + \frac{\mu}{2}\,\sum_{j=1}^p ( w_j)^2 + \lambda\,\sum_{j=1}^p \vert w_j\vert\, \Big\},
\label{appendix:eq:en2}
\end{equation}
the closed-form solution is explicitly given by the proximal operator $\text{prox}_{\frac{\mu}{2}\,\Vert\cdot\Vert_2^{\,2} + \lambda\,\Vert\cdot\Vert_1}$ that is in particular the composition of $\text{prox}_{\frac{\mu}{2}\,\Vert\cdot\Vert_2^{\,2}}$ and $\text{prox}_{\lambda\,\Vert\cdot\Vert_1}$ \citep[Theo. 4]{yu2013}, i.e.
\[
\text{prox}_{\frac{\mu}{2}\,\Vert\cdot\Vert_2^{\,2} + \lambda\,\Vert\cdot\Vert_1}(\mb c)  = \text{prox}_{\frac{\mu}{2}\,\Vert\cdot\Vert_2^{\,2}} \circ \text{prox}_{\lambda\,\Vert\cdot\Vert_1}(\mb c)\, .
\]
Both proximal operators $\text{prox}_{\lambda\,\Vert\cdot\Vert_1}$ and $\text{prox}_{\frac{\mu}{2}\,\Vert\cdot\Vert_2^{\,2}}$ are known \citep{bach2012}, respectively being:
\[
\begin{aligned}
& \text{prox}_{\lambda\,\Vert\cdot\Vert_1}(\mb c) = \Big( \text{sgn}(c_j)\,\big(\vert c_j\vert - \lambda\big)_+ \Big)_{j=1:p}, \\
& \text{prox}_{\frac{\mu}{2}\,\Vert\cdot\Vert_2^{\,2}}(\mb c) = \frac{1}{1+\mu}\,\mb c,
\end{aligned}
\]
where $\text{sgn}(\cdot)\,(\vert \cdot \vert - \lambda)_+$ is the soft-thresholding operator. Eventually, the coordinates of the solution are then:
\begin{equation}
\text{prox}_{\frac{\mu}{2}\,\Vert\cdot\Vert_2^{\,2} + \lambda\,\Vert\cdot\Vert_1}(\mb c) = \Big(\frac{1}{1+\mu} \, \text{sgn}(c_j)\,\big(\vert c_j\vert - \lambda\big)_+\Big)_{j=1:p}\, ,
\label{appendix:eq:prox}
\end{equation}
which correspond to the normalized soft-thresholding operator applied to the vector $\mb c = \tr{\mb X_c}\mbg\xi_c$.

We use the solution~(\ref{appendix:eq:prox}) of the Elastic Net problem~(\ref{appendix:eq:en2}), where $\lambda = \lambda_s$ and $\mu$ is chosen so that the solution has a unitary norm, to find a candidate point and then the solution (by convexity) of the dual problem~(\ref{appendix:eq:en}).

Finally, we have reformulated the problem defining the sparse PLS as a least squares problem with an Elastic Net penalty and we have shown that the solution of this problem is the (normalized) soft-thresholding operator.

\subsection{Adaptive penalty}

When considering an adaptive penalty, the optimization problem associated to the sparse PLS can be similarly rewritten as:
\begin{equation}
\left\{
\begin{aligned}
& \underset{\mb w\in\RR^p}\argmin \ \Big\{ \frac{1}{2} \Vert \mb c - \mb w \Vert_2^{\,2} + \sum_{j=1}^p \lambda_j\,\vert w_j\vert\, \Big\} , \\
& \Vert \mb w \Vert_2 = 1\, ,
\end{aligned}\right.
\label{appendix:eq:spls3}
\end{equation}
with the penalty constant $\lambda_j = \lambda\,\gamma^j$ (c.f. main text). By a similar reasoning (continuity and convexity), it is possible to use Lagrange multiplier to resolve the problem~(\ref{appendix:eq:spls3}).

In order to explicitly derive the solution, we will use the proximal operator that is solution of the following problem:
\begin{equation}
\underset{\mb w\in\RR^p}\argmin \ \Big\{ \frac{1}{2} \, \Vert \mb c - \mb w \Vert_2^{\,2} + \frac{\mu}{2}\,\sum_{j=1}^p (w_j)^2 + \sum_{j=1}^p \lambda_j\, \vert w_j\vert\, \Big\},
\label{appendix:eq:en3}
\end{equation}
with $\mu>0$.

If $f_1(\mb w) = \sum_j \lambda_j \vert w_j\vert$, it can be shown that the solution of the problem~(\ref{appendix:eq:prox0}) when considering $f=f_1$ is given by:
\[
\text{prox}_{f_1}(\mb c) = \Big(\text{sgn}(c_j)\,\big(\vert c_j\vert - \lambda_j\big)_+\Big)_{j=1:p},
\]
because the subgradient of $f_1$ is given by $\nabla^s f_1(\mb w) = \big( \lambda_j\,\text{sgn}(w_j) \big)_{j=1}^p$  \citep{eksioglu2011}.

The link between subgradient and proximal operator is described in \cite{bach2012}. In particular, $\mb w^* = \text{prox}_f(\mb c)$ if and only if $\mb c - \mb w^* \in \partial f(\mb w^*)$ for any couple $(\mb c,\mb w^*) \in \RR^p \times \RR^p$, where $\partial f(\mb w^*)$ is the subdifferential of $f$ at point $\mb w^*$, i.e. the set of all subgradients $\nabla^s f(\mb w^*)$ of $f$ at point  $\mb w^*$. If $f$ is differentiable in $\mb w^*$, then the only subgradient is the gradient $\nabla f(\mb w^*)$.

The proximal operator corresponding to the function $f_2(\mb w) = \frac{\mu}{2} \sum_j  (w_j)^2$ is known (c.f. previously):
\[
\text{prox}_{f_2}(\mb c) = \frac{1}{1+\mu}\,\mb c.
\]

Eventually, thanks to theorem 4 in \cite{yu2013}, the solution of problem~(\ref{appendix:eq:en3}) is explicitly defined as the combination of $\text{prox}_{f_1}$ and $\text{prox}_{f_2}$:
\[
\text{prox}_{f_1+f_2}(\mb c)  = \text{prox}_{f_2} \circ \text{prox}_{f_1}(\mb c)\, .
\]
for any $\mb c\in\RR^p$. Thus, the solution of the problem~(\ref{appendix:eq:en3}) is given by:
\[
\text{prox}_{f_1+f_2}(\mb c) = \Big(\frac{1}{1+\mu}\ \text{sgn}(c_j)\,\big(\vert c_j\vert - \lambda_j\big)_+\Big)_{j=1:p},
\]
with $\mb c=\tr{\mb X_c}\mbg\xi_c$.

We choose $\mu$ so that the norm of the solution is unitary to find a candidate point and thus the solution (by convexity) of the adaptive problem~(\ref{appendix:eq:spls3}).

\section{Conditions for stability selection}\label{supp:sec:cond_stab_sel}

The result by \cite{meinshausen2010} regarding the expected number of wrongly selected variables is derived for $\Lambda\subset\RR^+$ under two conditions: $(i)$ assuming that the indicators $\big(\II_{\{j\in \hat{S}_\ell\}}\big)_{j\in S_0^{\, c}}$ are exchangeable for any $\ell\in\Lambda$. $(ii)$ The original procedure of selection is not worst than random guessing. The first assumptions assumes that the considered method does not ``prefer'' to select some covariates rather than some other in the set of the non-pertinent predictors. This hypothesis seems reasonable in our SPLS framework. The second one is verified according to the results on our simulations (c.f. section~3). Moreover, in the method we consider, the grid of hyper-parameters lies in $(\RR^+)^3$, however the parameter that truly influences the sparsity of the estimation is the parameter $\lambda_s\in\RR^+$. Therefore, the sparse PLS appears to be a reasonable framework to apply the concept of stability selection.

\section{Comparison with state-of-the-art approaches}
\label{supp:sec:comparison}


In the literature, other methodologies have been proposed to adapt (sparse) PLS for binary classification. We detail here different approaches based on (sparse) PLS and GLMs, especially regarding the potential issues raised by the combination of two optimization frameworks.

\paragraph{PLS and GLMs.} To overcome the convergence issue in the IRLS algorithm, \cite{marx1996} proposed to solve the weighted least square problem at each IRLS step with a PLS regression, i.e. $\hat{\mbg\beta}^{(t+1)}$ is computed by weighted PLS regression of the pseudo-response $\mbg \xi^{(t)}$ onto the predictors $\mb X$. However, such iterative scheme does not correspond to the optimization of an objective function. Hence, the convergence of the procedure cannot be guaranteed and the potential solution is not clearly defined.

Alternatively, \cite{wang1999} and \cite{nguyen2002} proposed to achieve the dimension reduction before the logistic regression. Their algorithm use the PLS regression as a preliminary compression step. The components $[\mb t_k]_{k=1}^K$ in the subspace of dimension $K$ are then used in the logistic regression instead of the predictors. Therefore, the IRLS algorithm does not deal with high dimensional data (as $K < p$). In this context, the PLS algorithm treats the discrete response as continuous. Such approach seems counter-intuitive as it neglects the definition of PLS to resolve a linear regression problem and it ignores the inherent heteroskedastic context. This algorithm is called PLS-log in the following. It can be noted that \cite{nguyen2002} or \cite{boulesteix2004} also proposed to use discriminant analysis as a classifier after the PLS step. This method, known as PLS-DA, is not directly linked to the GLM framework but we cite it as an alternative for classification with PLS-based approaches. It can be noted that \cite{barker2003} proposed a slightly different implementation of PLS-DA, which is however equivalent to \citeauthor{boulesteix2004}'s approach in the binary response case, since they both rely on equivalent univariate response PLS algorithms \citep{dejong1993, boulesteix2007}.

Then, \cite{ding2005} proposed the GPLS method. They introduced a modification in \citeauthor{marx1996}'s algorithm based on the Firth procedure \citep{firth1993}, in order to avoid the non-convergence and the potential infinite parameter estimation in logistic regression. However, this approach is also characterized by the absence of an explicit optimization criterion. Eventually, as introduced previously, \cite{fort2005} proposed to integrate the dimension reduction PLS step after a Ridge regularized IRLS algorithm. We presented the adaptation of such methodology in the context of sparse PLS in the previous section.

\paragraph{Sparse PLS and GLMs.} More recently, based on the SPLS algorithm by \cite{chun2010}, \cite{chung2010} presented two different approaches. The first one, called SGPLS, is a direct extension from the GPLS algorithm by \cite{ding2005}. It solves the successive weighted least square problems of IRLS using a sparse PLS regression, with the idea that variable selection reduces the model complexity and helps to overwhelm numerical singularities. Unfortunately, our simulations will show that convergence issues remain. Indeed, the use of SPLS does not resolve the issue link to the absence of an associated optimization problem. The second approach is a generalization of the PLS-log algorithm and uses sparse PLS to reduce the dimension before running the logistic regression on the SPLS components. This method will be called SPLS-log.In both case, i.e. in SGPLS and SPLS-log , the iterative optimization in the IRLS algorithm or modified IRLS algorithm does depend on the number $K$ of components and on the sparsity parameter $\lambda_s$. Thus, the convergence of the algorithm is potentially affected by the choice of the hyper-parameters.

Eventually, we cite the SPLS-DA method developed by \citep{chung2010} or \cite{lecao2011}. Generalizing the approach from \cite{boulesteix2004}, they used sparse PLS as a preliminary dimension reduction step before a discriminant analysis. In the binary response case, thanks to the equivalence between \cite{boulesteix2004} and \cite{barker2003} works, the sparse extension of \citeauthor{barker2003}' PLS-DA for binary classification corresponds to the work of \cite{chung2010} or \cite{lecao2011}. A disadvantage of sparse PLS-DA approaches is that, in the multi-group classification case, they both rely on multivariate response sparse PLS algorithms, which do not admit a closed-form solution. On the contrary, our approach uses a univariate response sparse PLS algorithm (which admits a closed-form solution, c.f. main text) in both binary and multi-group classifications, being computationally efficient in both cases.

\section{Performance evaluation}\label{supp:sec:imp}

In order to assess the performance of our method, we compare it to other state-of-the-art approaches taking into account sparsity and/or performing compression. We eventually use a ``reference'' method, called GLMNET \citep{friedman2010}, that performs variable selection, by solving the GLM likelihood maximization penalized by $\ell_1$ norm penalty for selection and $\ell_2$ norm penalty for regularization, also known as the Elastic Net approach \citep{zou2005}. Computations were performed using the software environment for statistics \texttt{R}. The GPLS approach used in our computation comes from the archive of the former \texttt{R}-package \texttt{gpls}, the methods logit-PLS and PLS-DA from the package \texttt{plsgenomics}, SGPLS, SPLS-log and SPLS-DA from the package \texttt{spls}, GLMNET from the package \texttt{glmnet}.

\section{Complements on the simulation study}
\label{supp:sec:comp_simu}

\subsection{Simulation design}\label{supp:subsec:simu}

We consider a predictor matrix $\mb X$ of dimension $n\times p$, with $n=100$ fixed, and $p=100, 500, 1000, 2000$, so that we examine low and high dimensional models.  To simulate redundancy within predictors, $\mb X$ is partitioned into $k^*$ blocks (10 or 50 in practice) denoted by $\mathcal{G}_k$ for block $k$. Then for each predictor $j \in\mathcal{G}_k$, $X_{ij}$ is generated depending on a latent variable $H_k$ as $X_{ij} = H_{ik} + F_{ij}$, with $H_{ik} \sim \mathcal{N}(0,\sigma_H^2)$  and some noise $F_{ij} \sim \mathcal{N}(0,\sigma_F^2)$. The correlation between the blocks is regulated by $\sigma_H^2$, the higher $\sigma_H^2$ the less dependency. In the following we consider $\sigma_H / \sigma_F = 2$ or 1/3.

The true vector of predictor coefficients $\mbg\beta^*$ is structured according to the blocks $\mathcal{G}_k$ in $\mb X$. Actually, $\ell^*$ blocks in $\mbg\beta^*$ are randomly chosen  among the $k^*$ ones to be associated with non null coefficients (with $\ell^* = 1$ or $k^*/2$).  All coefficients within the $\ell^*$ designated blocks are constant (with value 1/2). In our model, the relevant predictors contributing to the response will be those with non zero coefficient, and our purpose will be to retrieve them via selection. The response variable $Y_i$ for observation $i$ is sampled as a Bernoulli variable, with parameter $\pi_i^*$ that follows a logistic model: $\pi_i^* = \text{logit}^{-1}(\tr{\mb x_{i}}\mbg\beta^*)$.

For our method, the parameter values that are tuned by cross-validation are the following: the number of components $K$ varies from 1 to 10, candidate values for the Ridge parameter $\lambda_R$ in RIRLS are 31 points that are $\text{log}_{10}$-linearly spaced in the range $[10^{-2};10^3]$, candidate values for the sparse parameter $\lambda_s$ are 10 points that are linearly spaced in the range $[0.05;0.95]$. Other SPLS approaches (SGPLS and SPLS-log) only depend on hyper-parameters $(\lambda_s, K)$ for which candidate values are the same as for our method. Regarding GLMNET, we let the procedure chooses by itself the grid of hyper-parameters, as recommended by the authors in the documentation.

\subsection{Additional simulation results}\label{supp:add_res_simu}

\paragraph{Convergence.}
Tab.~\ref{supp:tab:simu:conv_cv} summarized the convergence of the different methods (logit-SPLS, SGPLS and SPLS-log) during the cross-validation procedure (including the tuning of $K$) on the simulations, depending on the number of predictors $p$. Our approach logit-SPLS always converges on contrary to other SPLS approaches for logistic regression.

\setlength{\tabcolsep}{10pt}
\begin{table}[!t]
\processtable{Averaged percentage of model that converged during cross-validation tuning of hyper-parameters for different values of $p$.\label{supp:tab:simu:conv_cv}} {\begin{tabular}{@{}lcccc@{}}\toprule Method & $p=100$ & $p=500$ & $p=1000$ & $p=2000$ \\\midrule
sgpls & $37$ & $34$ & $33$ & $33$ \\
spls-log & $44$ & $67$ & $71$ & $74$ \\
\textbf{logit-spls} & \textbf{100} & \textbf{100} & \textbf{100} & \textbf{100} \\\botrule
\end{tabular}}{}
\end{table}
\setlength{\tabcolsep}{3pt}

Tab.~\ref{supp:tab:simu:conv} summarized the convergence of the different methods on the simulations, when fitting the model after tuning the hyper-parameters (including $K$) by cross-validation, depending on the number $p$ of predictors. Again, our approach logit-SPLS always converges on contrary to other SPLS approaches for logistic regression.

\setlength{\tabcolsep}{10pt}
\begin{table}[!t]
\processtable{Averaged percentage of model fitting that converged over 75 simulations for different values of $p$. Hyper-parameters are tuned by cross-validation. \label{supp:tab:simu:conv}} {\begin{tabular}{@{}lcccc@{}}\toprule Method & $p=100$ & $p=500$ & $p=1000$ & $p=2000$ \\\midrule
gpls & $66$ & $59$ & $61$ & $56$ \\
sgpls & $33$ & $23$ & $23$ & $23$ \\
spls-log & $84$ & $52$ & $39$ & $32$ \\
\textbf{logit-spls} & \textbf{100} & \textbf{100} & \textbf{100} & \textbf{100} \\\botrule
\end{tabular}}{}
\end{table}
\setlength{\tabcolsep}{3pt}

In addition, Tab.~\ref{supp:tab:simu:conv_cv} shows the percentage of convergence for the different SPLS approaches across cross-validation repeated runs. We see a similar pattern as in Tab.~\ref{supp:tab:simu:conv}, only our method logit-SPLS almost certainly converge.

\setlength{\tabcolsep}{10pt}
\begin{table}[!t]
\processtable{Averaged precentage of runs that converged across repeated cross-validations (tuning of all hyper-parameters, including $K$).\label{supp:tab:simu:conv_cv}} {\begin{tabular}{@{}lcccc@{}}\toprule Method & $p=100$ & $p=500$ & $p=1000$ & $p=2000$ \\\midrule
sgpls & $37$ & $34$ & $33$ & $33$ \\
spls-log & $44$ & $67$ & $71$ & $74$ \\
\textbf{logit-spls} & \textbf{100} & \textbf{100} & \textbf{100} & \textbf{100} \\\botrule
\end{tabular}}{}
\end{table}
\setlength{\tabcolsep}{3pt}

\paragraph{Cross-validation stability.}

\begin{figure}[!tpb]
\centering
\includegraphics[width=0.99\linewidth]{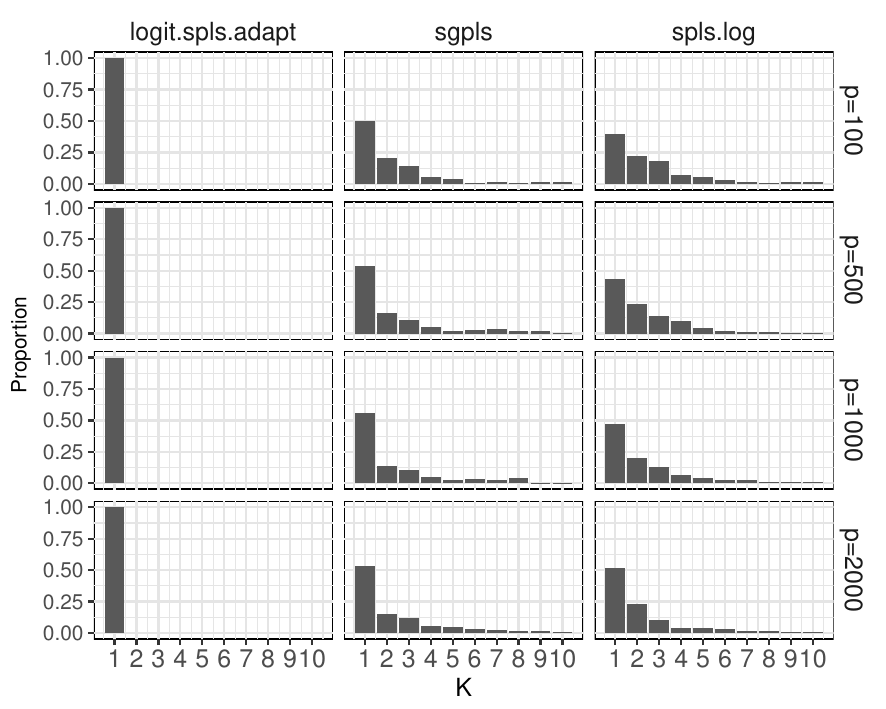}
\caption{\small Values chosen for $K$ by cross-validation over repetitions for the different SPLS approaches for different values of $p$ (with $n=100$) on simulated data.}
\label{supp:fig:simu:cv_K}
\end{figure}

Fig.~\ref{supp:fig:simu:cv_K} illustrates the stability of the cross-validation procedure for the different SPLS approaches regarding the number of components. Our approach logit-SPLS always chooses $K=1$, while other SPLS approaches mostly returns $K=1$. A first comment can be made on the stability of the cross-validation procedure. Our approach is also more stable regarding the choice of $K$ compared to other SPLS methods. A second comment is that, as explained in the manuscript, the stability of the cross-validation is directly linked to convergence of the method (c.f. Tab.~\ref{supp:tab:simu:conv}). Our method always converges on our simulations and is thus more stable regarding cross-validation than other SPLS approaches that do not converge most of the time and are less stable when tuning hyper-parameters. In addition, based on these results, we decided to set $K=1$ and only tune the sparsity parameter $\lambda_s$ and the Ridge parameter $\lambda_R$ when evaluating the performance of the different approaches (Tab.~3 in the manuscript) to save computation time.

\paragraph{Prediction and selection}

Tabs.~\ref{supp:tab:simu:pred_sel1},~\ref{supp:tab:simu:pred_sel2} and~\ref{supp:tab:simu:pred_sel3} collects the results regarding performance in prediction and selection (sensitivity, specificity, accuracy) for the different approaches compared in the simulation study, and for data respectively simulated with $p=100,500,1000$. These results are consistent with the case $p=2000$ presented in the manuscript. In details, approaches that combines compression and variable selection (sparse PLS) achieve better prediction performance than compression only (PLS) or selection only (GLMNET) approaches. Regarding selection, sparse PLS is generally better in term of selection sensitivity (true positive rate) compared to GLMNET, which is too conservative. However, our approach logit-SPLS seems to select less false positives compared to other SPLS approaches, since the specificity is higher for a similar accuracy level.

\setlength{\tabcolsep}{8.5pt}
\begin{table}[!t]
\processtable{Prediction error and selection sensitivity/specificity (if relevant) when $p=100$, for non-sparse or sparse approaches (delimited by the line).\label{supp:tab:simu:pred_sel1}} {\begin{tabular}{@{}lcccc@{}}
\toprule
\multirow{2}{*}{Method} & Prediction & Selection & Selection & Selection \\
& error & sensitivity & specificity & accuracy \\
\midrule
gpls & $0.51 \pm 0.30$ & / & / & / \\
pls-da & $0.22 \pm 0.08$ & / & / & / \\
logit-pls & $0.20 \pm 0.08$ & / & / & / \\
\midrule
glmnet & $0.17 \pm 0.08$ & $0.77$ & $0.76$ & 0.71 \\
\textbf{logit-spls} & $\mbg{0.14 \pm 0.07}$ & $\mbg{0.78}$ & $\mbg{0.86}$ & $\mbg{0.83}$ \\
sgpls & $0.14 \pm 0.07$ & $0.86$ & $0.77$ & $0.83$ \\
spls-da & $0.15 \pm 0.07$ & $0.88$ & $0.75$ & $0.83$ \\
spls-log & $0.12 \pm 0.07$ & $0.87$ & $0.75$ & $0.82$\\
\botrule
\end{tabular}}{}
\end{table}
\setlength{\tabcolsep}{3pt}

\setlength{\tabcolsep}{8.5pt}
\begin{table}[!t]
\processtable{Prediction error and selection sensitivity/specificity (if relevant) when $p=500$, for non-sparse or sparse approaches (delimited by the line).\label{supp:tab:simu:pred_sel2}} {\begin{tabular}{@{}lcccc@{}}
\toprule
\multirow{2}{*}{Method} & Prediction & Selection & Selection & Selection \\
& error & sensitivity & specificity & accuracy \\
\midrule
gpls & $0.47 \pm 0.31$ & / & / & / \\
pls-da & $0.22 \pm 0.08$ & / & / & / \\
logit-pls & $0.19 \pm 0.07$ & / & / & / \\
\midrule
glmnet & $0.18 \pm 0.07$ & $0.49$ & $0.93$ & 0.74 \\
\textbf{logit-spls} & $\mbg{0.13 \pm 0.07}$ & $\mbg{0.69}$ & $\mbg{0.85}$ & $\mbg{0.80}$ \\
sgpls & $0.12 \pm 0.06$ & $0.81$ & $0.76$ & $0.81$ \\
spls-da & $0.14 \pm 0.07$ & $0.82$ & $0.75$ & $0.81$ \\
spls-log & $0.13 \pm 0.06$ & $0.83$ & $0.77$ & $0.81$\\
\botrule
\end{tabular}}{}
\end{table}
\setlength{\tabcolsep}{3pt}

\setlength{\tabcolsep}{8.5pt}
\begin{table}[!t]
\processtable{Prediction error and selection sensitivity/specificity (if relevant) when $p=1000$, for non-sparse or sparse approaches (delimited by the line).\label{supp:tab:simu:pred_sel3}} {\begin{tabular}{@{}lcccc@{}}
\toprule
\multirow{2}{*}{Method} & Prediction & Selection & Selection & Selection \\
& error & sensitivity & specificity & accuracy \\
\midrule
gpls & $0.48 \pm 0.31$ & / & / & / \\
pls-da & $0.21 \pm 0.07$ & / & / & / \\
logit-pls & $0.18 \pm 0.07$ & / & / & / \\
\midrule
glmnet & $0.17 \pm 0.06$ & $0.37$ & $0.96$ & 0.74 \\
\textbf{logit-spls} & $\mbg{0.13 \pm 0.06}$ & $\mbg{0.66}$ & $\mbg{0.85}$ & $\mbg{0.80}$ \\
sgpls & $0.12 \pm 0.06$ & $0.80$ & $0.77$ & $0.81$ \\
spls-da & $0.13 \pm 0.06$ & $0.82$ & $0.75$ & $0.81$ \\
spls-log & $0.13 \pm 0.06$ & $0.83$ & $0.75$ & $0.81$\\
\botrule
\end{tabular}}{}
\end{table}
\setlength{\tabcolsep}{3pt}

\paragraph{Computation time.}

Tab.~\ref{supp:tab:time_cv} shows the averaged computation time for the cross-validation runs of the different approaches on simulated data where $n=100$ and $p=100,500,1000,2000$. Each run was performed on the cluster grid of the LBBE, equipped with standard multi-core CPU with frequency between 2 and  2.5 GHz. For each method, each cross-validation runs did used a single core of a single CPU for two reason: $(i)$ we did perform massive simultaneous runs on the cluster). $(ii)$ It was a fare basis for comparison, because the different packages that we used propose different degrees of parallelization in their implementation. It is important to note that our approach logit-SPLS can run on multi-core architecture, which improves the results presented below.

GLMNET is the most efficient method because its implementation relies on \texttt{fortran} and \texttt{C} codes, interfaced with \texttt{R}. SPLS-log is also quite efficient (less than a minute in all cases). Indeed, it uses the \texttt{glm} function from \texttt{R} that is encoded in \texttt{C}. However, as mentioned earlier and in the paper, this function did encounter convergence issues in many cases. Our method logit-SPLS is slower since the cross-validation takes between $\sim 1$ min. (when $p=100$) and $\sim 11$ min. (when $p=2000$) in average. We can make two comments here: $(i)$ our approach needs to calibrate an additional hyper-parameter $\lambda_R$, however this additional cost is reasonable (a few minutes). $(ii)$ The fast convergence of our approach ensures a lower computation time compared to the SGPLS approach, despite the additional hyper-parameter.

In addition, it can be noted that we are currently working on a \texttt{C++} implementation of our algorithm, which is expected to speed up the computations compared to the \texttt{R} implementation.

\setlength{\tabcolsep}{10pt}
\begin{table}[!t]
\processtable{Averaged computation time (in seconds) of cross-validation runs on a single-core of a standard CPU, when considering simulated data where $n=100$ and $p=100,500,1000,2000$. \label{supp:tab:time_cv}} {\begin{tabular}{@{}lcccc@{}}\toprule Method & $p=100$ & $p=500$ & $p=1000$ & $p=2000$ \\
\midrule
glmnet & 4.69 & 4.85 & 5.39 & 6.59 \\
\textbf{logit-spls} & \mbg{72.98} & \mbg{223.13} & \mbg{452.21} & \mbg{706.86} \\
sgpls & 79.41 & 284.62 & 541.86 & 1103.32 \\
spls-log & 3.63 & 11.17 & 20.74 & 37.30 \\
\botrule
\end{tabular}}{}
\end{table}
\setlength{\tabcolsep}{3pt}

Eventually, Tab.~\ref{supp:tab:time_fit} presents the averaged computation time to fit a single model for the different approaches on simulated data where $n=100$ and $p=100,500,1000,2000$. Each run was performed on the cluster grid of the LBBE, equipped with standard multi-core CPU with frequency between 2 and  2.5 GHz. For each method, each model fitting runs did used a single core of a single CPU.

All methods are computationally efficient to fit a single model, except for SGPLS. The non-convergence of this approach requires that the algorithm iterates further, until the limit set by the users. It can be noted that the cost of additional iterations in the case of SPLS-log is counter-balanced by the efficient use of the \texttt{glm} function. However, it does not guarantee its convergence (c.f. previously).

\setlength{\tabcolsep}{10pt}
\begin{table}[!t]
\processtable{Averaged computation time (in seconds) of a single fit run on a single-core of a standard CPU, when considering simulated data where $n=100$ and $p=100,500,1000,2000$. \label{supp:tab:time_fit}} {\begin{tabular}{@{}lcccc@{}}\toprule Method & $p=100$ & $p=500$ & $p=1000$ & $p=2000$ \\
\midrule
glmnet & 0.01 & 0.03 & 0.06 & 0.11 \\
\textbf{logit-spls} & \mbg{0.05} & \mbg{0.17} & \mbg{0.35} & \mbg{0.60} \\
sgpls & 0.89 & 3.70 & 7.86 & 17.92 \\
spls-log & 0.03 & 0.09 & 0.19 & 0.40 \\
\botrule
\end{tabular}}{}
\end{table}
\setlength{\tabcolsep}{3pt}

\section{Complements on the breast cancer data analysis}
\label{supp:sec:comp_data1}
\subsection{Computation details}\label{supp:subsec:comp_data1}

We applied the methods GLMNET, logit-PLS, logit-SPLS (adaptive or not), SGPLS and SPLS-log to our data set. We fit each model over a hundred resamplings, where observations are randomly split into training and test sets with a 70\%/30\% ratio. \text{}{For the prediction task,} on each resampling, the parameter values of each method are tuned by 10-fold cross-validation on the training set, respecting the following grid (for our method logit-SPLS) $K \in\{1,\dots,8\}$, candidate values for the Ridge parameter $\lambda_R$ in RIRLS are 31 points that are $\text{log}_{10}$-linearly spaced in the range $[10^{-2};10^3]$, candidate values for the sparse parameter $\lambda_s$ are 10 points that are linearly spaced in the range $[0.05;0.95]$. Other SPLS approaches (SGPLS and SPLS-log) only depend on hyper-parameters $(\lambda_s, K)$ for which candidate values are the same as for our method. Regarding GLMNET, we let the procedure chooses by itself the grid of hyper-parameters, as recommended by the authors in the documentation.

\subsection{Stability selection}\label{supp:subsec:stab_sel_data1}

\paragraph{Hyper-parameter grid.}
In the study of the stability selection on the breast cancer data set, regarding our approach logit-SPLS, as a basis, we use the same grid $\Lambda$ of candidates values for $(\lambda_s, \lambda_R, K)$ as in the cross-validation case (c.f. section \ref{supp:subsec:comp_data1}). As stated in the manuscript, the grid is then reduced to control the false positive expected number. For other SPLS approaches, we apply the same procedure, based on the grid $(\lambda_s,K)$. Regarding GLMNET, the grid of candidate values for the penalty parameter is chosen by the procedure itself, but then we apply the same framework to extract the set of stable selected variables (as detailed in the manuscript, section \ref{subsec:stab_sel}).

\begin{figure}[!tpb]
\centering
\includegraphics[width=0.99\linewidth]{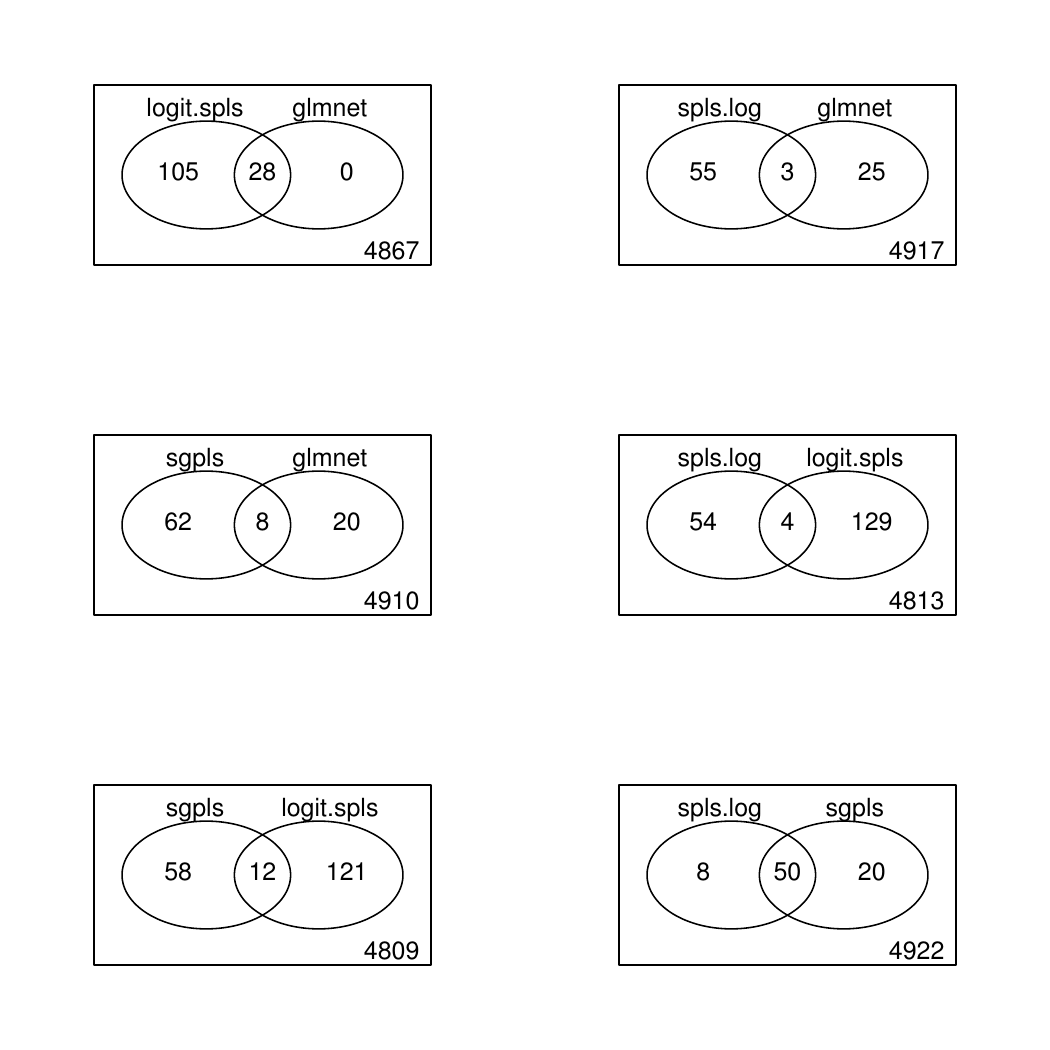}
\caption{\small Overlap between the genes selected by the different methods thanks to the stability selection procedure, when taking a threshold $\pi_\text{thr}=0.75.$}
\label{supp:fig:stab_sel}
\end{figure}

\paragraph{Selected genes.}
The overlap between the genes selected by the different approaches based on the stability selection procedure (for a threshold $\pi_\text{thr} = 0.75)$ are given in Fig.~\ref{supp:fig:stab_sel}. We can make two comments: $(i)$ The 28 genes selected by GLMNET are all retrieved by our approach logit-SPLS (over 133 selected genes). In addition, genes with higher selection score (i.e. maximum estimated probability to be selected) are the same between the two methods (c.f. Tabs.~\ref{supp:tab:glmnet_sel} and~\ref{supp:tab:logit_spls_sel}). Thus, the selection procedure based on our logit-SPLS method is consistent with our baseline GLMNET. $(ii)$ Genes selected by other SPLS-based approaches (SPLS-log and SGPLS) are not consistent with the ones selected by GLMNET nor by logit-SPLS. On the contrary, they select 50 common genes over respectively 58 and 70 selected genes for SPLS-log and SGPLS. However, the reliability of these results is questioned because of the non-convergence of these two methods (c.f. Tab.~\ref{tab:data:res} in the manuscript). It can be noted that similar observations (consistency between logit-SPLS and GLMNET, SPLS-log and SGPLS are different) can be made for other level of probability threshold $\pi_\text{thr}$.

Tabs.~\ref{supp:tab:glmnet_sel} and~\ref{supp:tab:logit_spls_sel} give the list of genes that were selected respectively by GLMNET and logit-SPLS thanks to the stability selection procedure applied to the breast cancer data set (in particular to the 5000 most differentially expressed genes between the two conditions relapse or not). Genes are identified by their ProbeID on Affymetrix U133-Plus2.0 chip \citep[c.f.][]{guedj2012}. Gene identification (Symbool, Entrezid and Name) is recovered thanks to the \texttt{annotate} and \texttt{hgu133plus2.db} \texttt{R}-package, that are available on Bioconductor (\texttt{https://www.bioconductor.org}). Some ProbeID were not identified and correspond to blank line. On contrary, other ProbeID seem to correspond to two genes and are present twice.

\setlength{\tabcolsep}{10pt}
\begin{table*}[!t]
\processtable{List of the 28 genes selected by GLMNET thanks to the stability selection procedure (at threshold $\pi_\text{thr}=0.75)$ on the breast cancer data set. Genes are sorted by selection score (maximum estimated probability to be selected). Genes are identified by their ProbeID on Affymetrix U133-Plus2.0 chip. \label{supp:tab:glmnet_sel}} {\begin{tabular}{@{}llll@{}r}
\toprule
PROBEID & SYMBOL & ENTREZID & GENE NAME & Selection score \\ 
\midrule
217048\_at &  &  &  & 0.99 \\ 
  1553561\_at & TAS2R50 & 259296 & taste 2 receptor member 50 & 0.97 \\ 
  233227\_at & KIAA1109 & 84162 & KIAA1109 & 0.97 \\ 
  218307\_at & RSAD1 & 55316 & radical S-adenosyl methionine domain containing 1 & 0.97 \\ 
  241034\_at & GLS & 2744 & glutaminase & 0.97 \\ 
  211870\_s\_at & PCDHA3 & 56145 & protocadherin alpha 3 & 0.95 \\ 
  211870\_s\_at & PCDHA2 & 56146 & protocadherin alpha 2 & 0.95 \\ 
  1561665\_at & LOC100421171 & 100421171 & thyroid hormone receptor interactor 11 pseudogene & 0.95 \\ 
  216738\_at &  &  &  & 0.92 \\ 
  236899\_at &  &  &  & 0.91 \\ 
  227240\_at & NGEF & 25791 & neuronal guanine nucleotide exchange factor & 0.91 \\ 
  234739\_at &  &  &  & 0.91 \\ 
  1560522\_at & DLGAP1-AS3 & 201477 & DLGAP1 antisense RNA 3 & 0.89 \\ 
  1554988\_at & SLC9C2 & 284525 & solute carrier family 9 member C2 (putative) & 0.86 \\ 
  217360\_x\_at & IGHA1 & 3493 & immunoglobulin heavy constant alpha 1 & 0.85 \\ 
  217360\_x\_at & IGHG1 & 3500 & immunoglobulin heavy constant gamma 1 (G1m marker) & 0.85 \\ 
  217360\_x\_at & IGHG3 & 3502 & immunoglobulin heavy constant gamma 3 (G3m marker) & 0.85 \\ 
  217360\_x\_at & IGHM & 3507 & immunoglobulin heavy constant mu & 0.85 \\ 
  217360\_x\_at & IGHV4-31 & 28396 & immunoglobulin heavy variable 4-31 & 0.85 \\ 
  229485\_x\_at & SHISA3 & 152573 & shisa family member 3 & 0.85 \\ 
  239052\_at &  &  &  & 0.85 \\ 
  242870\_at &  &  &  & 0.84 \\ 
  228776\_at & GJC1 & 10052 & gap junction protein gamma 1 & 0.83 \\ 
  244849\_at & SEMA3A & 10371 & semaphorin 3A & 0.83 \\ 
  217391\_x\_at &  &  &  & 0.82 \\ 
  229215\_at & ASCL2 & 430 & achaete-scute family bHLH transcription factor 2 & 0.82 \\ 
  235945\_at &  &  &  & 0.82 \\ 
  1554708\_s\_at & SPATA6L & 55064 & spermatogenesis associated 6 like & 0.79 \\ 
  208777\_s\_at & PSMD11 & 5717 & proteasome 26S subunit, non-ATPase 11 & 0.79 \\ 
  213651\_at & INPP5J & 27124 & inositol polyphosphate-5-phosphatase J & 0.78 \\ 
  225792\_at & HOOK1 & 51361 & hook microtubule tethering protein 1 & 0.78 \\ 
  1570136\_at &  &  &  & 0.76 \\ 
  1560692\_at & VSTM2A-OT1 & 285878 & VSTM2A overlapping transcript 1 & 0.75 \\ 
  1560692\_at & VSTM2A & 222008 & V-set and transmembrane domain containing 2A & 0.75 \\ 
\botrule
\end{tabular}}{}
\end{table*}
\setlength{\tabcolsep}{3pt}

\setlength{\tabcolsep}{10pt}
\begin{table*}[!t]
\processtable{List of the top 50 genes (over 133) selected by logit-SPLS thanks to the stability selection procedure (at threshold $\pi_\text{thr}=0.75)$ on the breast cancer data set. Genes sorted by selection score (maximum estimated probability to be selected). Genes are identified by their ProbeID on Affymetrix U133-Plus2.0 chip. \label{supp:tab:logit_spls_sel}} {\begin{tabular}{@{}llll@{}r}
\toprule
PROBEID & SYMBOL & ENTREZID & GENE NAME & Selection score \\ 
\midrule
1553561\_at & TAS2R50 & 259296 & taste 2 receptor member 50 & 1.00 \\ 
  218307\_at & RSAD1 & 55316 & radical S-adenosyl methionine domain containing 1 & 1.00 \\ 
  1560522\_at & DLGAP1-AS3 & 201477 & DLGAP1 antisense RNA 3 & 0.99 \\ 
  217048\_at &  &  &  & 0.99 \\ 
  211870\_s\_at & PCDHA3 & 56145 & protocadherin alpha 3 & 0.99 \\ 
  211870\_s\_at & PCDHA2 & 56146 & protocadherin alpha 2 & 0.99 \\ 
  233227\_at & KIAA1109 & 84162 & KIAA1109 & 0.99 \\ 
  220098\_at & HYDIN & 54768 & HYDIN, axonemal central pair apparatus protein & 0.98 \\ 
  220098\_at & HYDIN2 & 100288805 & HYDIN2, axonemal central pair apparatus protein (pseudogene) & 0.98 \\ 
  234739\_at &  &  &  & 0.98 \\ 
  1561665\_at & LOC100421171 & 100421171 & thyroid hormone receptor interactor 11 pseudogene & 0.97 \\ 
  216738\_at &  &  &  & 0.97 \\ 
  241034\_at & GLS & 2744 & glutaminase & 0.97 \\ 
  227240\_at & NGEF & 25791 & neuronal guanine nucleotide exchange factor & 0.97 \\ 
  1554988\_at & SLC9C2 & 284525 & solute carrier family 9 member C2 (putative) & 0.95 \\ 
  1560692\_at & VSTM2A-OT1 & 285878 & VSTM2A overlapping transcript 1 & 0.95 \\ 
  1560692\_at & VSTM2A & 222008 & V-set and transmembrane domain containing 2A & 0.95 \\ 
  217360\_x\_at & IGHA1 & 3493 & immunoglobulin heavy constant alpha 1 & 0.95 \\ 
  217360\_x\_at & IGHG1 & 3500 & immunoglobulin heavy constant gamma 1 (G1m marker) & 0.95 \\ 
  217360\_x\_at & IGHG3 & 3502 & immunoglobulin heavy constant gamma 3 (G3m marker) & 0.95 \\ 
  217360\_x\_at & IGHM & 3507 & immunoglobulin heavy constant mu & 0.95 \\ 
  217360\_x\_at & IGHV4-31 & 28396 & immunoglobulin heavy variable 4-31 & 0.95 \\ 
  236899\_at &  &  &  & 0.95 \\ 
  239052\_at &  &  &  & 0.95 \\ 
  242870\_at &  &  &  & 0.95 \\ 
  228507\_at & PDE3A & 5139 & phosphodiesterase 3A & 0.95 \\ 
  229081\_at & SLC25A13 & 10165 & solute carrier family 25 member 13 & 0.95 \\ 
  1562030\_at & LOC284898 & 284898 & uncharacterized LOC284898 & 0.94 \\ 
  227379\_at & MBOAT1 & 154141 & membrane bound O-acyltransferase domain containing 1 & 0.94 \\ 
  225792\_at & HOOK1 & 51361 & hook microtubule tethering protein 1 & 0.93 \\ 
  234792\_x\_at & IGHA1 & 3493 & immunoglobulin heavy constant alpha 1 & 0.93 \\ 
  234792\_x\_at & IGHV4-31 & 28396 & immunoglobulin heavy variable 4-31 & 0.93 \\ 
  244849\_at & SEMA3A & 10371 & semaphorin 3A & 0.93 \\ 
  217697\_at &  &  &  & 0.93 \\ 
  1556937\_at &  &  &  & 0.92 \\ 
  1569126\_at & CCNC & 892 & cyclin C & 0.92 \\ 
  228776\_at & GJC1 & 10052 & gap junction protein gamma 1 & 0.92 \\ 
  229485\_x\_at & SHISA3 & 152573 & shisa family member 3 & 0.92 \\ 
  232920\_at & KIAA1656 & 85371 & KIAA1656 protein & 0.92 \\ 
  232920\_at & CCDC157 & 550631 & coiled-coil domain containing 157 & 0.92 \\ 
  1563057\_at &  &  &  & 0.91 \\ 
  1568666\_at & PLIN5 & 440503 & perilipin 5 & 0.91 \\ 
  1570116\_at &  &  &  & 0.91 \\ 
  238824\_at & RPS29 & 6235 & ribosomal protein S29 & 0.91 \\ 
  243583\_at &  &  &  & 0.91 \\ 
  206349\_at & LGI1 & 9211 & leucine rich glioma inactivated 1 & 0.91 \\ 
  211064\_at & ZNF493 & 284443 & zinc finger protein 493 & 0.91 \\ 
  231913\_s\_at & BRCC3 & 79184 & BRCA1/BRCA2-containing complex subunit 3 & 0.91 \\ 
  1570136\_at &  &  &  & 0.89 \\ 
  206202\_at & MEOX2 & 4223 & mesenchyme homeobox 2 & 0.89 \\ 
\botrule
\end{tabular}}{}
\end{table*}
\setlength{\tabcolsep}{3pt}

\section{Sparse PLS for multi-group classification}\label{supp:sec:multinom_spls}

We generalize our approach to a multi-categorical response. This problem is known as multinomial logistic regression or polytomous regression \citep{mccullagh1989} and will be called multinomial sparse PLS in the sequel.

\subsection{Multinomial logistic regression}

The response $y_i$ takes its values in a discrete set $\{0,\hdots,G\}$ corresponding to $G+1$ groups or classes of observations. The associated variable $Y_i$ ($i=1,\hdots,n$) follows a multi-categorical distribution where $\PP(Y_{ij} = g\,\vert\,\mb x_i) = 
\pi_{ig}$ for any class $g$. Based on a direct generalization of the logistic model, a class of reference is set (generally the class 0) and 
for each class $g\ne 0$, the probability $\pi_{ig}$ that $Y_i = g$ depends on a linear combination of predictor such as:
\begin{equation}
\log\left(\frac{\pi_{ig}}{\pi_{i0}}\right) = \tr{\mb z_i}\mbg\beta_g,
\label{supp:eq:multinom}
\end{equation}
with a specific vector of coefficient $\mbg\beta_g\in\RR^{p+1}$ for each class $g=1,\hdots,G$. Indeed, the probabilities $(\pi_{ig})_{g=1:G}$ determine the probability $\pi_{i0}$ since $\msum_{g=0}^G\, \pi_{ig} = 1$. A column of 1s is added in the matrix $\mb Z$ to incorporate the intercept in the linear combination $\tr{\mb z_i}\mbg\beta_g$. The log-likelihood can be be explicitly formulated:
\begin{equation}
\log\like\big(\mbg\beta\big) = \sum_{i=1}^n \left\{ \sum_{g=1}^G y_{ig}\,\tr{\mb z_i}\mbg\beta_g - \log\left(1+\msum_{g=1}^G \exp(\tr{\mb z_i}\mbg\beta_g)\right)\right\},
\label{part1:chap:log_spls:eq:loglike_multinom}
\end{equation}
where the binary variable $y_{ig} = \II_{\{y_i = g\}}$ indicates the class of the observation $i$ ($\II_{\{\mathcal{A}\}}$ is the indicator function valued in $\{0,1\}$, indicating if the statement $\mathcal{A}$ is true (1) or false (0).)

It is possible to rearrange the data in order to formulate a vectorized version of the loss~(\ref{part1:chap:log_spls:eq:loglike_multinom}), and express the multinomial logistic regression as a logistic regression of a binary response $\mathcal{Y} \in\{0,1\}^{n\,G}$ against a matrix of rearranged covariates $\mathcal{Z}\in\RR^{n\,G \times (p+1)\,G}$. The response vector $\mathcal{Y}$ of length $n\,G$ is 
defined as follows:
\[
\mathcal{Y} = \tr{\Big((y_{1g})_{g=1:G}, \hdots, (y_{ig})_{g=1:G}, \hdots, (y_{ng})_{g=1:G} \Big)},
\]
where $y_{ig} = \II_{\{y_i = g\}}$ as previously mentioned. The new covariate matrix $\mathcal{Z}$ of dimension $n\,G \times 
(p+1)\,G$ is defined by blocks as:
\[
\mathcal{Z} = \tr{\left[\tr{\mathcal{Z}_1}, \hdots, \tr{\mathcal{Z}_i}, \hdots, \tr{\mathcal{Z}_n} \right]},
\]
where each block $i$ is constructed by $G$ diagonal repetitions of the row $\tilde{\mb x}_i$ from the original covariate matrix $\mb Z$, i.e.
\[
\mathcal{Z}_i = \left. \begin{pmatrix} 1\ x_{i1}\ \hdots\ x_{ip} &  & 0 \\
& \ddots & \\
0 &  & 1\ x_{i1}\ \hdots\ x_{ip} \\
\end{pmatrix} \ \ \ \ \right\} \text{$G$ repeats of row $\tr{\mb z_i}$}.
\]
The coefficient vectors $\mbg\beta_g\in\RR^{p+1}$ (for $g=1,\hdots,G$) are also reorganized in the vector $\mb B\in\RR^{(p+1)\,G}$ as:
\[
\mb B = \tr{\Big((\beta_{0g})_{g=1:G}, \hdots, (\beta_{jg})_{g=1:G}, \hdots, (\beta_{pg})_{g=1:G} \Big)},
\]
where $(\beta_{jg})_{j=0:p}$ are the coordinates of $\mbg\beta_g$, so that the response $\mathcal{Y}$ depends on the linear combination 
$\mathcal{Z}\,\mb B$.

Thanks to this reformulation, it is possible to adapt the Ridge IRLS algorithm to estimate the coefficients $\mb B$ and infer the probabilities $\pi_{ig}$ that observations $y_i$ belongs to the class $g$. The algorithm that we call MRIRLS is detailed in \cite{fort2005a}.

\subsection{Multinomial SPLS}\label{supp:subsec:multinom_spls}

The vectorized formulation of the MIRLS algorithm allows to use our SPLS-based dimension reduction approach. As in the binary case, the MIRLS algorithm (penalized by Ridge) produces a continuous pseudo-response (at the convergence) that is suitable for the sparse PLS regression. Thus, our approach, called multinomial-SPLS, directly extends our algorithm logit-SPLS to the multinomial logistic regression. It estimates the linear coefficients $\mb B$ by sparse PLS, processing compression and variable selection simultaneously. Then, these estimated coefficients are used to get an estimation of the probabilities $\pi_{ig}$. Our procedure is directly inspired from the approach by \cite{fort2005a} that extended the algorithm logit-PLS \citep{fort2005} to the multi-categorical cases.

In this context, the SPLS step considers: $i)$ the pseudo-response $\mbg\xi\in\RR^{n\,G}$ constructed from the reformulated response $\mathcal{Y}$, $ii)$ the centered version $\mathcal{X}_c$ of the modified covariate matrix $\mathcal{X}$ defined by:
\[
\mathcal{X} = \tr{\left[\tr{\mathcal{X}_1}, \hdots, \tr{\mathcal{X}_i}, \hdots, \tr{\mathcal{X}_n} \right]},
\]
where each block $i$ is constructed by $G$ diagonal repetitions of the row $\tilde{\mb x}_i$ from the original covariate matrix $\mb X$, i.e.
\[
\mathcal{X}_i = \left. \begin{pmatrix} x_{i1}\ \hdots\ x_{ip} &  & 0 \\
& \ddots & \\
0 &  & x_{i1}\ \hdots\ x_{ip} \\
\end{pmatrix} \ \ \ \ \right\} \text{$G$ repeats of row $\tr{\mb x_i}$}.
\]
It corresponds to the matrix $\mathcal{Z}$ where the terms 1 corresponding to the intercept have been removed. Thus, the coefficients $(\beta_{0g})_{g=1:G}$ are estimated afterward. These coefficients are ultimately used to compute the class membership probabilities for each observation, following the model~(\ref{supp:eq:multinom}). In prediction task, an observation is assigned to the class with the highest predicted probability.

At this point, we mention that the error rate that we consider in this case (especially for tuning of hyper-parameters by $V$-fold cross-validation, with $V=5$ or 10) is the standard error rate, i.e. the proportion of overall mismatches, that previously used in the \texttt{plsgenomics} \texttt{R}-package for multi-class PLS classification.

\subsection{SPLS components}

Since the sparse PLS is applied on the modified covariate matrix $\mathcal{X}\in\RR^{n\,G \times p\,G}$, the constructed SPLS components represent the matrix $\mathcal{X}$ in a lower dimensional subspace, and not the original matrix $\mb X$. However, it is possible to obtain a low dimensional representation of the original covariates. Indeed, thanks to the construction of the matrix $\mathcal{X}$, the SPLS weight vectors $\mb w_k\in\RR^{p\,G}$ are partitioned as follows:
\[
\mb w_k = \tr{\Big((w_{j1}^k)_{j=1:p}, \hdots, (w_{jg}^k)_{j=1:p}, \hdots, (w_{jG}^k)_{j=1:p} \Big)},
\]
for $k=1,\hdots,K$. Thus when multiplying the original predictor matrix $\mb X$ by the weights matrix $\big[(w_{jg}^k)_{j=1:p}\big]_{k=1:K} \in\RR^{p\times K}$, we obtain a representation of the observations in a lower dimensional space of dimension $K$, as a matrix $\mb T_g\in\RR^{n\times K}$. The matrix $\mb T_g$ represents the directions that discriminate the class $g$ versus the class reference 0.

\subsection{State-of-the-art}

It can be noted that \cite{ding2005} presented a version of the GPLS method suitable for multinomial logistic regression, i.e. the linear regression inside the iteration of the MIRLS algorithm are processed by weighted PLS regression. \cite{chung2010} introduced 
a similar algorithm based on sparse PLS (extension of the SGPLS algorithm). However, we used exclusively our multinomial SPLS algorithm in the data analysis. Indeed, based on the conclusions from the binary case, our approach showed better results regarding prediction performance on an experimental data set. Moreover, the dimension of the data is drastically increased because of the rearrangement since the number of observations becomes $n\,G$ and the number of covariates becomes $p\,G$. It is therefore necessary to account for the computational cost and to give priority to computationally efficient methods. In particular, thanks to the Ridge penalty, we showed that our approach converges quickly, hence reducing the time of computation.

\section{Complements on the single T cell data analysis}\label{supp:sec:comp_data2}

\subsection{Computation details}\label{supp:subsec:comp_data2}

On each resampling, the parameter values of each method are tuned by 10-fold cross-validation on the training set, respecting the following grid $K \in\{1,\dots,4\}$, candidate values for the Ridge parameter $\lambda_R$ in RIRLS are 10 points that are $\text{log}_{10}$-linearly spaced in the range $[10^{-2};10^3]$, candidate values for the sparse parameter $\lambda_s$ are 10 points that are linearly spaced in the range $[0.05;0.95]$.

\subsection{Additional results single cell data analysis}\label{supp:subsec:add_data2}

\begin{figure*}[!tpb]
\centering
\includegraphics[width=0.99\linewidth]{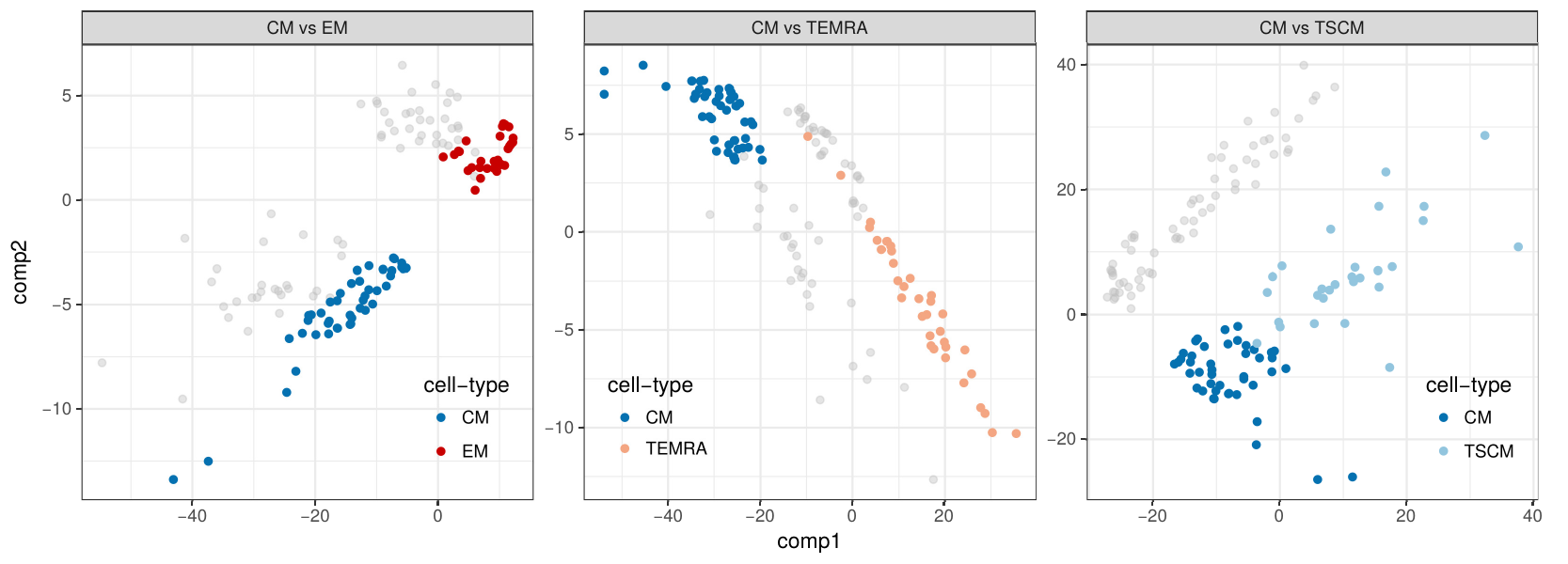}
\caption{\small Cell scores on the first two PLS components in the latent space that discriminate between the reference class (``CM'') and each other class separately (``EM'', ``TEMRA'' and ``TSCM'' respectively, from left to right). Restriction to the T cells in the training set before the first prediction step.}
\label{supp:fig:data:cell_type2}
\end{figure*}

\paragraph{Training in the first step of prediction.} The manual identification of cells is mainly based on the level of the CCR7 markers. The identified cells mostly correspond to the most extreme values of CCR7 level. The set of manually identified cells constitutes the training set for the first step of prediction based on multinomial sparse PLS. Fig.~\ref{supp:fig:data:cell_type2} illustrates the representation of the cells in the training set according to the first two PLS components. The distinction between the reference class (``CM'') and both classes from the group of ``Effector'' cells (``EM'' and ``TEMRA'') is clearly apparent in the latent subspace, since there is an important gap between the different groups of cells. It confirms that the cells in the training set correspond to the most extreme phenotypes that appear clearly different.

\paragraph{Genes selection by sparse PLS.}

The genes that are selected by the multinomial-SPLS during the second round of prediction (as explained in the manuscript) are the following: ``CCL4'', ``CCR7'', ``CST7'', ``GNLY'', ``GZMB'', ``KLRD1'', ``LTB'', ``S100A4''. These genes have been identified as genes involved in the phenotype (``Effector'' or ``Memory'') of T-cells \citep{wherry2007, willinger2005}. In particular, ``CCR7'' and ``LTB'' are associated to ``Memory'' cells, while ``CCL4'', ``CST7'', ``GNLY'', ``GZMB'' and ``KLRD1'' characterized ``Effector'' cells.

\end{document}